\documentclass[12pt,preprint]{aastex}
\slugcomment{Not to appear in Nonlearned J., 45.}

\shorttitle{Harmonic QPOs in BL Lac Object AO 0235+164}
\shortauthors{Liu et al.}

\begin{document}

\title{Harmonic QPOs and Thick Accretion Disk Oscillations in BL Lac
  Object AO 0235+164}

\author{F.K. Liu, G. Zhao \and Xue-Bing Wu}
\email{fkliu@bac.pku.edu.cn, wuxb@bac.pku.edu.cn}
\affil{Astronomy Department, Peking University, 100871 Beijing, China}

\begin{abstract}

     Periodic outbursts are observed in many AGNs and usually
     explained with a supermassive black hole binary (SMBHB)
     scenario. However, multiple periods are observed in some AGNs and
     cannot be explained with it. Here we analyze
     the periodicity of the radio light curves of AO 0235+164 at 
     multi-frequencies and report the discovery of six QPOs in integer
     ratio 1:2:3:4:5:6 of QPO frequencies, of which the second with
     period $P_2 = (5.46 \pm 0.47) \, {\rm yr}$ is the strongest.  We
     fit the radio light curves and show that the initial phases of
     six QPOs have zero or $\pi$ differences relative to each
     other. We suggest a harmonic relationship of QPOs. The
     centroid frequency, relative strength, harmonic relationship and
     relative initial phases of QPOs are independent of radio
     frequency. The harmonic QPOs are likely due to the quasi-periodic
     injection of plasma from an oscillating accretion disk into the
     jet. We estimate the supermassive black hole mass $M_{\rm BH}
     \simeq (4.72\pm 2.04) \times 10^8 M_\odot$ and the accretion rate
     $\dot{m}\simeq 0.007$.   With the knowledge of accretion disk, it
     implies that the inner region of accretion disk of AO 0235+164 
     is a radiatively inefficient accretion flow. The oscillation
     accretion is due to the p-mode oscillation of the thick disk
     probably excited by a SMBHB. The theoretical predications of
     fundamental oscillation frequency and the harmonics are well
     consistent with the observations. 
     Harmonic QPOs would be absent when the thick disk
     becomes geometrically thin due to the increase of accretion
     rate. We discuss the observations of AO 0235+164 basing on the
     SMBHB-thick disk oscillation scenario. 
     
\end{abstract}


\section{Introduction}

   Blazars, consisting of BL Lac objects and flat-spectrum radio
   quasars, are a subclass of AGNs whose relativistic jet is nearly
   aligned to the line of sight, and show extreme variabilities at from
   radio through optical and X-ray to high-energy $\gamma$-ray
   wavelengths \citep{antonucci93,urry95,wagner95,ulrich97}. Long term
   multi-wavelength monitorings show that the variabilities of blazars
   are very complex with large amplitude and on time-scale ranging
   from tens of minutes, hours and days to months and years at all
   wavelengths. The outbursts with time-scale of
   order of years or longer in some blazars are  periodic
   \citep{sill88,liu95,liu97,rai01,qian04} and of particular
   interest as it might connect to the presence of supermassive black
   hole binary (SMBHB) at center \citep{sill88,liu02,qian04,
   osto04}, whose ultimate coalescence would generate an outburst
   of gravitational wave radiation \citep{thorne76} and be the main
   target of a future space gravitational wave detector, Laser
   Interferometer Space Antenna (LISA) \citep{merritt05}. 

   SMBHBs in galactic nuclei are expected by the hierarchical galaxy
   formation model in the cold dark matter (CDM) cosmology
   \citep{kauffmann00} in which the present galaxies are the products of
   frequent galaxy minor mergers. Active SMBHB and final
   coalescence have been suggested to be the physical origin of
   the peculiar radio morphologies in some AGNs \citep[for a recent
   review, see][]{komossa03}, e.g. the helical morphology of radio
   jets 
   \citep{begelman80}, the interruption and recurrence of activity in 
   double-double radio galaxies \citep{liu03}, the X-shaped feature of
   winged radio sources \citep{merritt02,liu04}, the orbital motion of
   radio core \citep[e.g.][]{sudou03}. Since they were
   discovered in BL Lac object OJ287 \citep{sill88}, periodic 
   optical outbursts of blazars have been ascribed to the orbital 
   motion of SMBHB \citep{sill88,abraham99,valtaoja00,liu02,
   rieger04,lobanov05}. However, it is unclear how SMBHBs in AGNs 
   trigger the observed periodic outbursts if present. The proposed
   scenarios in literature include (1) direct impact of the secondary
   black hole against a standard thin accretion
   disk \citep[e.g.][]{lehto96}) or an inefficient accretion flow,
   e.g. advection dominated accretion flow \citep[ADAF;][]{liu02}, 
   (2) rotating helical jets due to the orbital motion of SMBHB
   \citep{katz97,villata98,osto04,rieger04}, (3) periodic change of
   jet orientation introduced by the Lens-Thirring precessing of a
   warped disk \citep{abraham99,lobanov05}. Although the detailed
   physical mechanisms in these models are different, all of them 
   suggest a common feature of single period in the light curves.

   One more complication to the investigation of the periodicity in
   blazars is from the observations of low frequency X-ray
   quasi-periodic oscillations (QPOs) in black hole X-ray  binaries
   \citep{has86,klis89,stroh96} and possibly in Sgr A$^*$ of our
   Galactic center \citep{torok05}, which are believed to be due 
   to the disk oscillations. Black hole X-ray binaries, so called
   micro-quasars, are physical analogue to AGNs and are powered
   by accretion of stellar black holes of mass about $10 {\rm
   M}_\odot$. As the variation time-scale of accretion disk around
   black holes is proportional to the mass of black hole and the
   formation of jet is coupled with accretion disk 
   \citep{gallo03,fend04,fend04a}, the low frequency QPO with typical
   frequency $\sim 1 {\rm Hz}$ in micro-quasars corresponds to a
   period of order of years or longer in a blazar system of black hole
   mass larger than $10^8 M_\odot$. Disk oscillations may cause a
   periodic change of accretion and the QPOs in X-ray black hole
   binary system \citep{rezzolla03}, which could lead to periodic jet
   emission due to the jet-disk coupling. 

   To understand the physical origin of periodic outbursts, it is
   essential to investigate in details the structure of periodic
   outbursts in both light curves and power density spectrum, and the 
   dependence of periodicity on the observational wavelengths. For
   example, the discovery of double peaks in the periodic optical
   outbursts of the BL Lac object OJ287 \citep{sill96} leads to the
   second version of the SMBHB model \citep{lehto96}, while the
   investigation of the relation of the optical outburst structure and
   the radio variations leads to the third version of the SMBHB
   scenario \citep{valtaoja00,liu02}. It is generally believed that
   the short-term variabilities of blazars are non-periodic
   \citep[e.g.][]{wagner95,ulrich97}, but the short-term periodic
   brightness flickering may have been found to superpose on the
   long-term 
   periodic variations in some blazars, e.g. OJ287 \citep{wu06} and
   3C66A \citep{lain99}. In order to give more constraints on the
   models for periodic outbursts in AGNs, recently we started a
   program to investigate the relationships of periodic major
   outbursts and minor events and the long- and short-term
   periodicities of a large sample of AGNs \citep{zhao05}. Here we
   report the results of multiple quasi-periodic oscillations (QPOs)
   and the harmonic resonant relationship 1:2:3:$\dots$ of QPO
   frequencies in BL Lac object AO 0235+164.

   BL Lac object AO 0235+164 with redshift $z=0.94$ is one of the most
   violently variable object whose periodicity of outbursts in optical
   and radio bands has been investigated and reported in literature
   \citep[e.g.][]{web88,web00,roy00,rai01,rai05}. \citet{roy00}
   analyzed the data of the University of Michigan Radio Astronomy
   Observatory (UMRAO) at frequencies 4.8, 8.0 and 14.5
   GHz from 1975 to 1999 and discovered a period 
   $\sim 5.8 \, {\rm yrs}$ at 8.0 and 14.5 GHz, which is consistent
   with the observations of a period $5.7 \, {\rm yrs}$ in the optical
   \citep{rai01,rai05}. \citet{osto04} interpreted the long term
   periodicity of major outbursts with a helical-jet model and 
   the minor events with rotation of the non-periodic inhomogeneity.
   However, the radio outbursts are double-peaked and the burst peaks
   do not appear as regularly as the period predicted
   \citep{rai01,rai05}, implying that the substructure of the outburst
   are complex. On the other hand, the multiple short periods though at
   much lower significant level of signal-to-noise and the poor
   identification of central frequency  have been claimed in
   literature. \citet{web88} first identified the three peaks with
   three periods $\sim 2.79 \, {\rm yrs}$, $\sim 1.58\, {\rm yrs}$,
   and $\sim 1.29 \, {\rm yrs}$ in 
   the Fourier power spectrum of the optical light curves.
   Subsequent observations marginally confirmed the periods with 
   2.7 and 1.2 years in optical \citep{web00} and with 1.8, 2.8 and
   3.7 years in radio wave-bands with the Discrete Correlation Function
   (DCF) analysis method \citep{rai01}. 

   Our work is based on the three radio databases of the University of
   Michigan Radio Astronomy Observatory (UMRAO), the National
   Radio Astronomy Observatory (NRAO), and the Mets\"ahovi 
   Observatory. In Sec.~\ref{database}, we start the investigations
   from analyzing the consistency of UMRAO's and NRAO's databases, by
   comparing the best-sampled observations at 8.0 GHz at UMRAO and
   those at 8.2 GHz at NRAO and then merging the observational data at
   these two 
   frequencies. The periodic analysis results for the combined light
   curves at 8 GHz and identification of  harmonic QPOs are given in
   Sec.~\ref{result8}. In Sec.~\ref{resmult}, we 
   investigate the dependence of QPO properties on radio frequency.
   After estimating the mass of central supermassive black hole (SMBH)
   and the accretion rate in Sec.~\ref{sec:bhmass}, we analytically
   discuss the p-mode oscillation of thick disk in
   Sec.~\ref{sec:diskosc}.  A SMBHB-disk oscillation scenario for
   multiple harmonic QPOs is suggested and discussed in
   Sec.~\ref{sec:pertbinary}.  Our discussions and conclusions are
   given in Sec.~\ref{concl}. We assume a flat cosmology with $H_0 =
   75 \, {\rm Km \; s^{-1} Mpc^{-1}}$, $\Lambda = 0$, and $q_0 = 0.5$  
   throughout the paper, leading to a source distance $d \simeq 4.4 
   Gpc$. 

%========================================================

\section{Databases and the consistency}
\label{database}

    UMRAO has been monitoring AO 0235+164 at 4.8, 8.0, and 14.5 GHz
    for about three decades \citep{all85,all99}, while NRAO's database
    contains the observational data at 2.5 and 8.2 GHz for about 20
    years \citep{fie87,wal91,laz01}. At frequencies 22 and 37 GHz,
    the Mets\"{a}hovi Observatory started collecting the flux data
    since 1980 \citep{ter98,ter04}. As the central frequencies of 8.0
    GHz and 8.2 GHz are almost the same and light curves
    at the two frequencies are best sampled in both databases of UMRAO 
    and of NRAO, we investigate the consistence of the observations and 
    construct a combined light curves, based on the data from both 
    databases. The observations at UMRAO last about $\sim 25$~years 
    with a relatively sparse data sampling ($\sim 0.1$ data points per
    day), while those at NRAO cover a shorter time of about 15 years with 
    relatively dense data sampling ($\sim 0.5$ data points per day). To 
    remove the effects of intra-day variabilities \citep{rom97} and 
    of the noises due to unevenly sampling, we 
    compose two quasi-simultaneous observational light curves 
    by averaging the observational data with a 10-days time-interval,
    which is much shorter than the variability time-scale (of order of 
    months) of major outbursts but give, respectively, about one and
    five observations in each bin at 8.0 GHz and 8.2 GHz to
    smooth the intra-day random variations.

%-----
   \begin{figure*}
   \plotone{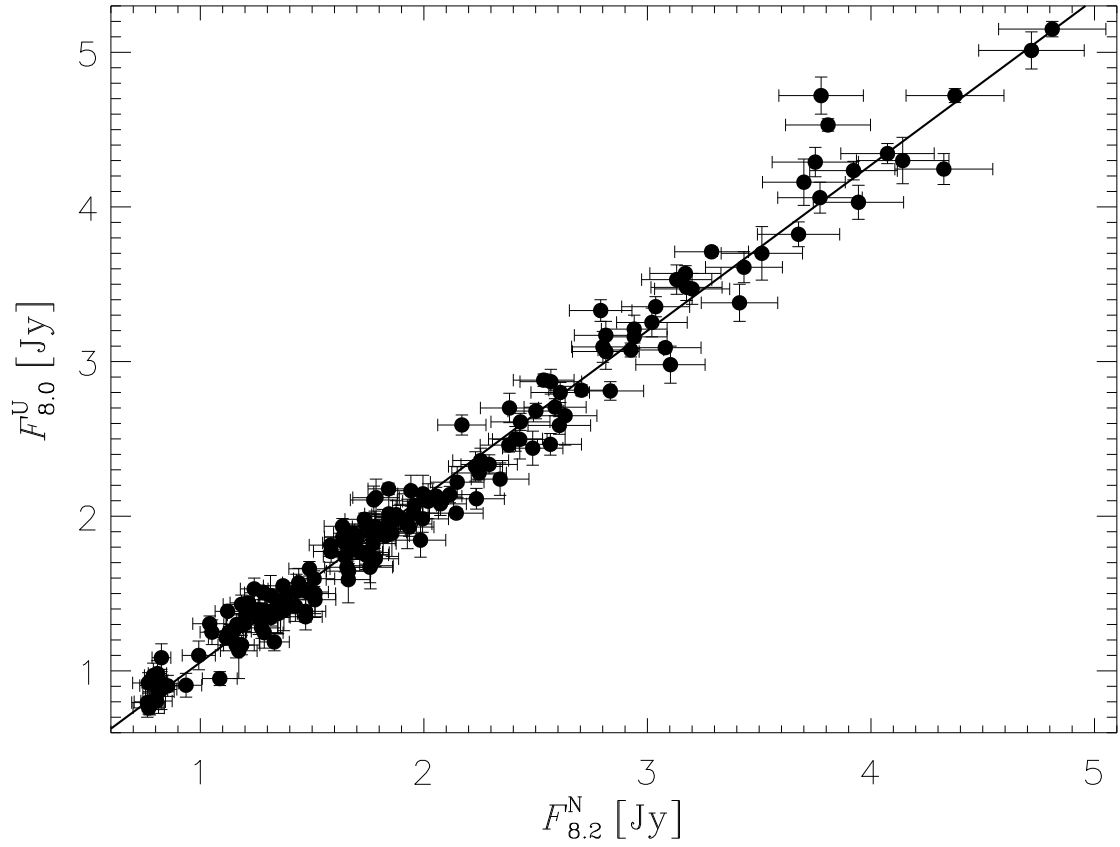}
   \caption{ 
      Correlations of the observations of UMRAO at 8.0 GHz 
     (y-axis) and of NRAO at 8.2 GHz  (x-axis).  The data are 10-days
      averaged and the solid line is a least square fit.}
     \label{correl}
   \end{figure*}
%------

    The results are shown in Fig.~\ref{correl}. The standard error in
    Fig.~\ref{correl} is the statistical error. A least square fit
    with a correlation coefficient $r=0.989$ gives 
%---
    \begin{equation}
      F_{8.0} = (-0.017 \pm 0.014) + (1.072 \pm 0.006) F_{8.2} ,
      \label{eq_cor}
    \end{equation}
%---
    where $F_{8.0}$ and $F_{8.2}$ are the radio flux at 8.0 GHz
    (UMRAO) and at 8.2 GHz (NRAO), respectively. The results
    show that the observations of UMRAO and of NRAO are
    consistent with each other with a linear relation but the slope is
    not exact unity. So we merged the
    observational data at 8.0 GHz  and at 8.2 GHz and obtain a
    combined light curve by converting NRAO's observations with the
    fitted relations Eq.~(\ref{eq_cor}). The combined radio light curve
    at 8 GHz is given in Fig.~\ref{fit8}, which shows that
    the combined light curves fully covers several major outbursts in 
    details although the combined data do not cover a significant
    longer time. In the following sections, we will give the periodic
    analysis 
    results and discussions, based on the combined light curve at 8
    GHz. In order to show how efficient and to what extent the
    combination of two databases can improve the results, we also
    analyze the radio light curves at 8.0 GHz and 8.2 GHz,
    respectively.

%===================

\section{Multiple QPOs in the combined 8 GHz light curve}
\label{result8}

\begin{figure*}
   \plotone{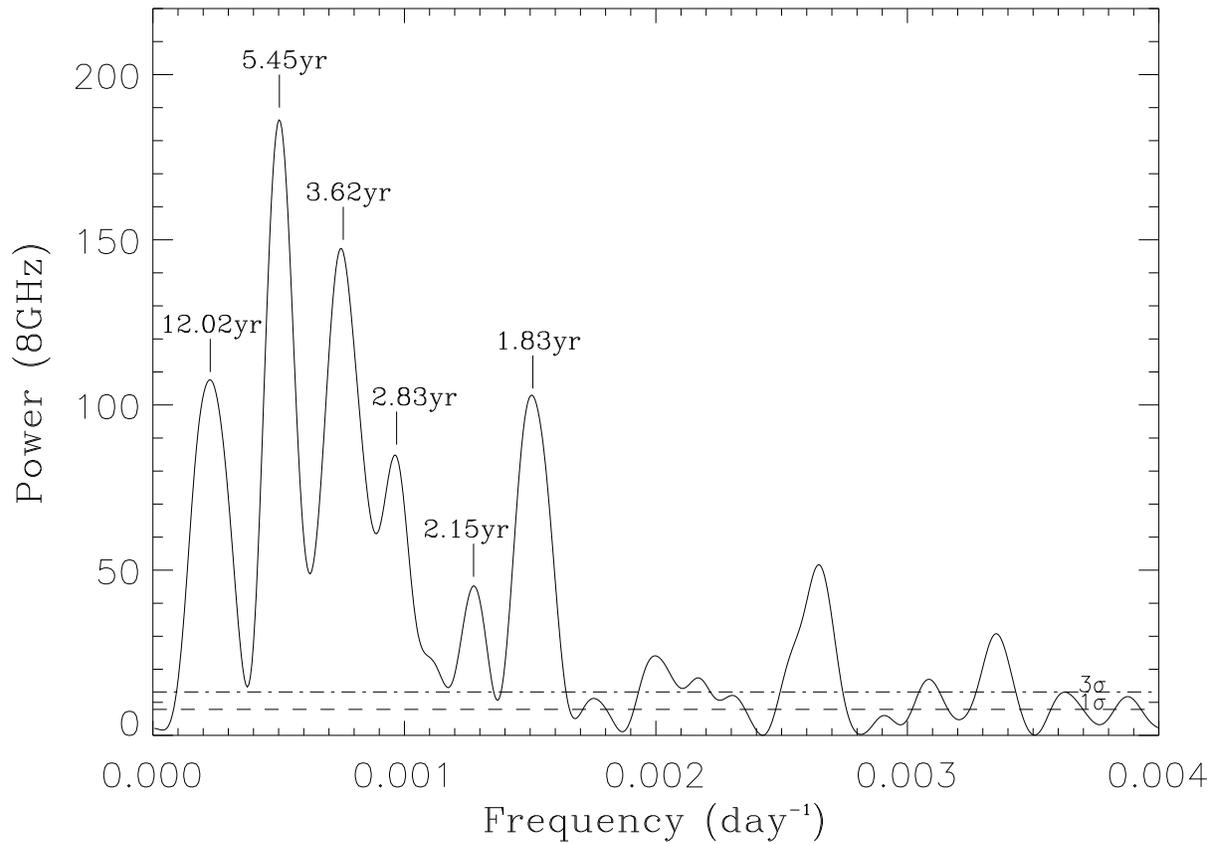}
   \caption{The power spectrum of the combined radio light curve at 8
     GHz. Fitted centroid periods of QPOs are indicated. The dashed
     and the dash-dotted lines are the $1 \sigma$ and $3 \sigma$
     significance levels, respectively.} 
   \label{0235l8c}
   \end{figure*}

    We analyze the periodicity of the combined light curve at 8 GHz
    by computing the Fourier power spectrum with a Lomb normalized
    periodo-gram \citep{lom76,sca82}, which is designed to properly
    treat the unevenly sampled data. The power spectrum is showed in
    Fig.~\ref{0235l8c}. There are several very prominent peaks in the
    power spectrum with very high signal-to-noise ratios. We identify
    a peak with a QPO if (1) it is at least 5 times higher than the
    nearby background noise; (2) it has a Gaussian profiles; (3) it is 
    also detected with other periodic analysis methods, e.g. the
    Jurkevich $\rm V_m^2$ method \citep{jur71} and the z-transform
    discrete correlation function (ZDCF) technique \citep{ale97};
    and/or (4) it correlates with other QPOs, e.g. with a harmonic
    relationship.

\begin{figure*}
   \plottwo{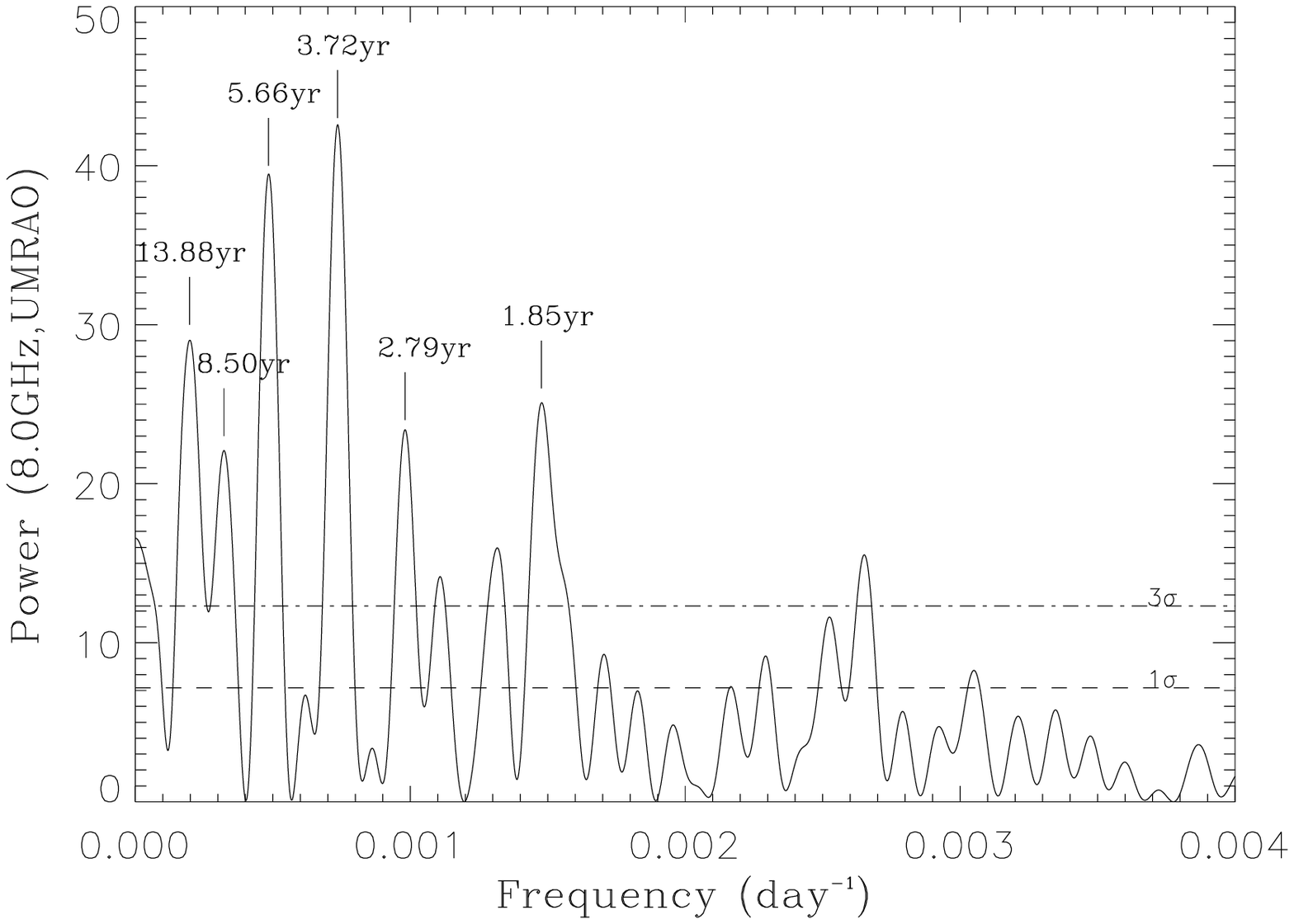}{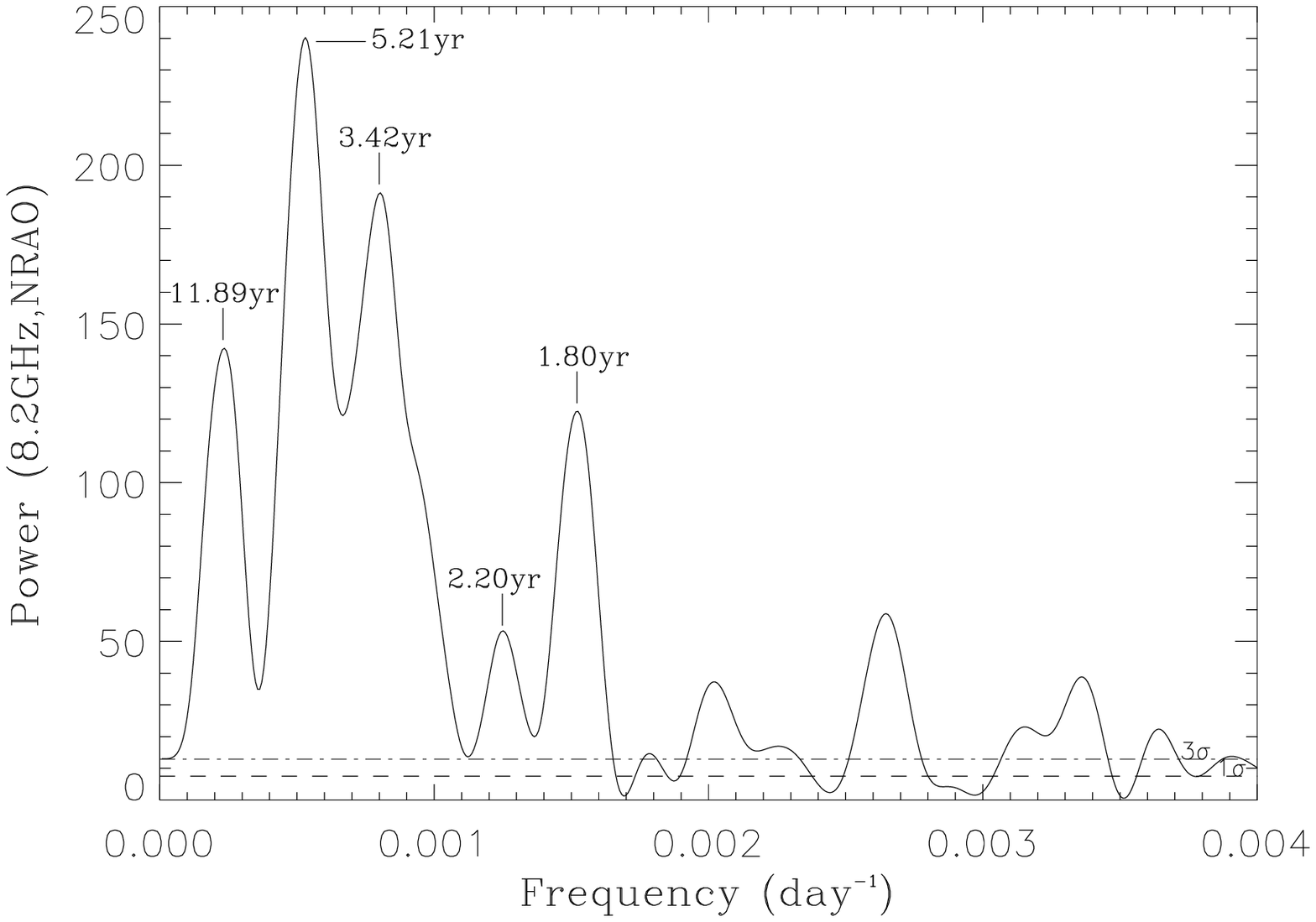}
   \caption{The power spectrum of the radio light curves at 8.0 GHz at
     UMRAO ({\it left}) and at 8.2 GHz at NRAO ({\it right}). The
     identified QPOs are indicated. The dashed and the dash-dotted
     lines are the $1 \sigma$ and $3 \sigma$ significance levels,
     respectively.} 
   \label{0235l8un}
   \end{figure*}

    Before discussing the periodic analysis results of the combined
    light curve, we firstly investigate the effects of the combination
    of the two databases. We analyze, respectively, with the Lomb method
    the light curves at 8.0 GHz at UMRAO and at 8.2 GHz at NRAO, and
    give the power spectra in Fig.~\ref{0235l8un}. 
    Fig.~\ref{0235l8c} and Fig.~\ref{0235l8un} show that all the
    peaks in Fig.~\ref{0235l8un} are present in Fig.~\ref{0235l8c} and 
    the combination of two databases significantly improves the
    signal-to-noise ratio of the power spectrum, which of the combined
    light curve is unprecedentedly high. This is because the radio
    light curve at 8.0 GHz has relatively sparse data points and misses
    the structure information of the major bursts, while the observations
    at 8.2 GHz span a shorter observational time interval and give
    only broad peaks in the power spectrum with small coherent quality
    factors. The second effect one can read from Fig.~\ref{0235l8c}
    and Fig.~\ref{0235l8un} is that though most of the peaks in the power
    spectrum of the combined radio light curve can be found in the
    power spectra of the radio light curve at 8.0 GHz and 8.2 GHz, the 
    central frequency of each peak is significantly different. A
    prominent peak of 8.50 years at the periodogram of radio light
    curve at 8.0 GHz is absent both at the power spectra of the
    combined radio light curve and of the light curve at 8.2
    GHz. Therefore, it is probably spurious because of the low
    signal-to-noise ratio due to the incomplete coverage of major
    outbursts in the 
    sparsely-sampled light curve at 8.0 GHz and to the loss of the
    burst structure information. To solve the problem, much more
    intensive observations are need.  

%--------------
%-----------------------
%---------------------

\begin{figure*}
 \plotone{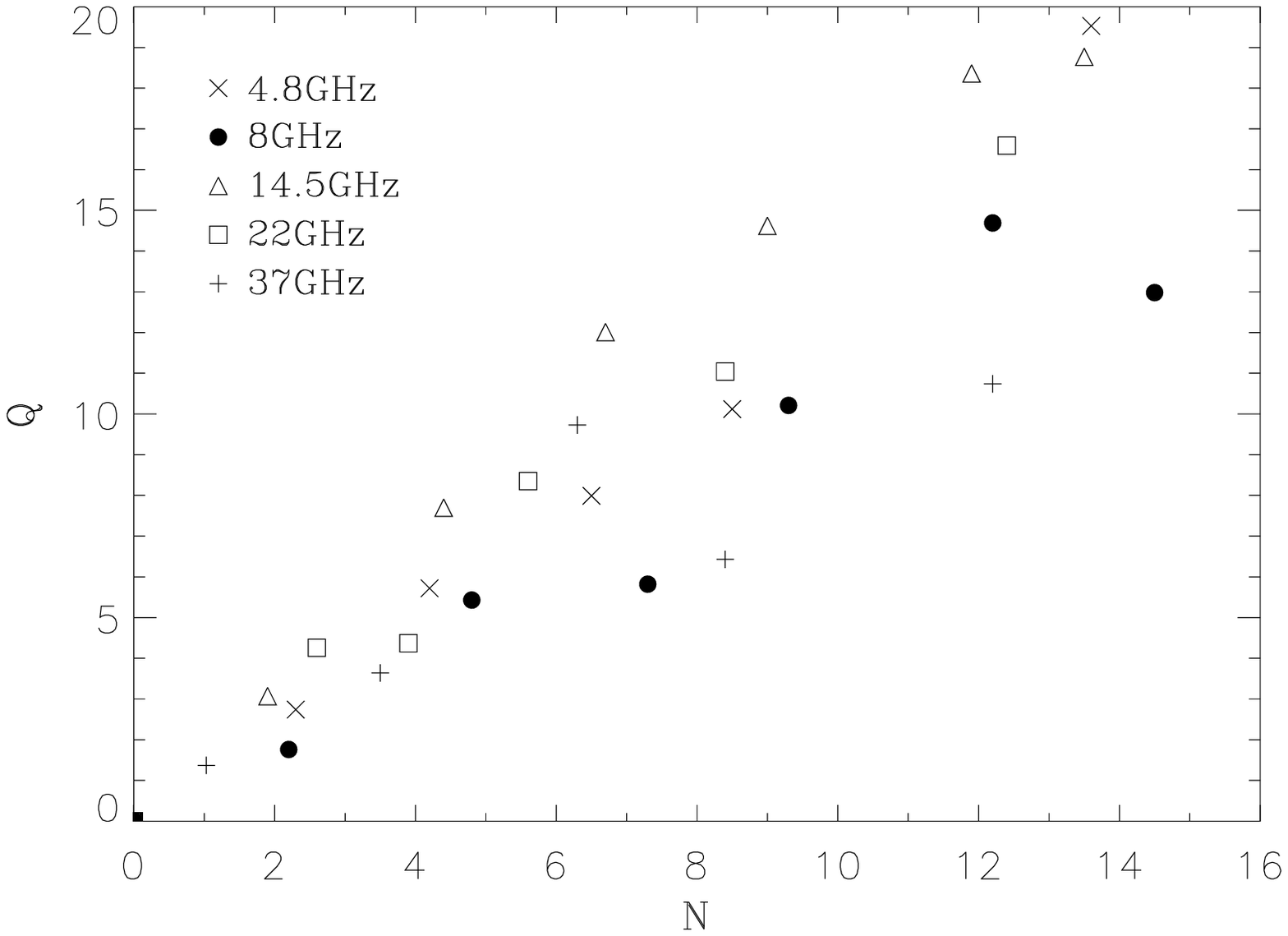}
   \caption{Quality factor Q {\it vs} repetition number N of QPOs in
     the combined radio light curves. The correlation of Q and 
     N implies that the Q-value is limited by the monitoring time
     of the object and much longer observations are needed to give
     stringent constraint on the intrinsic coherence of QPOs.}   
   \label{0235qn}
\end{figure*}

    The peaks in Fig.~\ref{0235l8c} are Gaussian, though some of them
    overlap with each other. Therefore, we fit each peak profile with a
    Gaussian function or, if necessary, fit several peak profiles together
    with Gaussian functions. The fitting results are given in
    Table~\ref{idf}. In Table~\ref{idf}, Col.~1 gives the observational
    radio frequency $\nu_{\rm obs}$, the averaged period of the
    fundamental QPO and their standard errors. We will discuss the
    definition of fundamental QPO and how to calculate the average
    period 
    later. The centroid frequency $\nu$ and the standard error
    $\sigma$ of  each peak are given in Col.~2 and indicated in
    Fig.~\ref{0235l8c}. The fitting goodness $\chi^2$ is presented in
    Col.~6. The standard error $\sigma$ is that of a fitting
    Gaussian function and is considered to contain all the effects,
    including random variations in the exact 
    quasi-periodic oscillation, the large and changing width of the
    outburst structure, the poor coverage of some outbursts, the random 
    variations in intensity, and in particular the intrinsic
    coherence of variability. To quantify 
    the coherence of the variabilities, we calculate the quality
    factor $Q=\nu/\rm HWHM$ and give it in Col.~5, where HWHM
    is the half width at half maximum of a peak. The quality factor
    $Q$ in Table~\ref{idf} increases with the QPO frequency $\nu$ and
    correlates with the repetitions of a QPO in the combined radio
    light curve, which is calculated with $N = \nu \Delta
    {T}$, where $\Delta{T} \simeq 26.3 \, {\rm yr}$ is the total
    observation time of the combined data sample. A strong
    correlation of the quality factor $Q$ and the repetition number
    $N$ is showed in Fig.~\ref{0235qn}, implying that the coherence
    factor Q-value and the quality of QPOs are limited by the duration
    of the monitoring program. Therefore, to understand the intrinsic
    coherence of a QPO, we have to monitor the object for a much
    longer time. 

%---
%---

    In the Lomb periodogram of the combined 8 GHz light curve, we
    identify six QPOs of frequencies (periods) $\nu_1 = 0.2279 \times 
    10^{-3} \, {\rm day}^{-1}$ ($P_1 = 12.02 \, {\rm yr}$), $\nu_2 = 
    0.5029 \times 10^{-3} \, {\rm day}^{-1}$ ($P_2 = 5.45 \, {\rm
    yr}$), $\nu_3 = 0.7568 \times 10^{-3} \, {\rm day}^{-1}$ ($P_3 =
    3.62 \, {\rm yr}$), $\nu_4 = 0.9691 \times 10^{-3} \, {\rm
    day}^{-1}$ ($P_4 = 2.83 \, {\rm yr}$), $\nu_5 = 1.275 \times
    10^{-3} \, {\rm day}^{-1}$ ($P_5 = 2.15 \, {\rm yr}$), and $\nu_6
    = 1.511 \times 10^{-3} \, {\rm day}^{-1}$ ($P_6 = 1.81 \, {\rm
    yr}$). The periods of the QPOs and the standard errors are given
    in Col.~4 of Table~\ref{idf}. 
    To estimate the relative contribution of each QPO signal to the
    total flux variations, we compute the relative root mean square
    ({\it rms}) in percentage and give the results in Col.~3 of
    Table~\ref{idf}. The results show that the QPO of period $P = 5.45
    \, {\rm yrs}$ is the strongest with $rms \sim 10$~per cent, though
    the other QPOs are also significant. We identify the frequency 
    of the strongest QPO with the fundamental one with 
    frequency $\nu_s$. 

\begin{figure*}
   \plotone{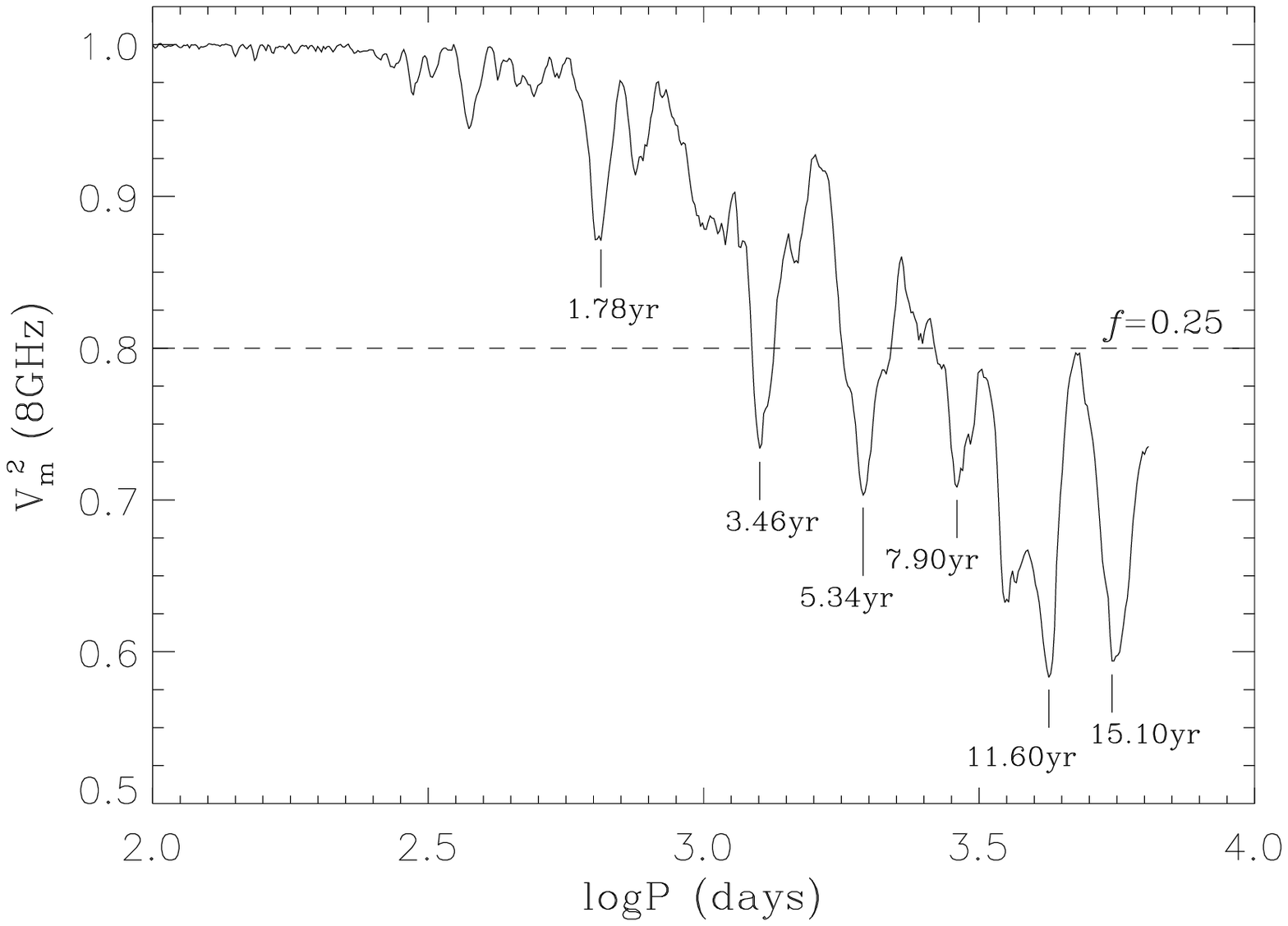}
   \caption{Normalized Jurkevich periodic analysis results of 
     the combined radio light curve at 8 GHz. The minima corresponding 
     to periods 1.78 yrs, 3.46 yrs, 5.34 yrs, 7.90 yrs and 11.60 yrs
     are prominent. A period is strong if $ f> 0.25$ with $f = (1-V_m^2)
     / V_m^2$}.  
   \label{0235j}
   \end{figure*}

    In order to confirm the detection of multiple QPOs, we now apply
    the least variance analysis introduced by 
    \citet{jur71} and the z-transform discrete correlation function
    (ZDCF) technique \citep{ale97} to the combined 8 GHz light curve.
    Jurkevich method has been successfully
    developed to search for periods in the optical light curves of
    some blazars \citep[e.g.][]{kid92,liu95,liu97}. The method 
    is based on the Expected Mean Square Deviation and specifically
    designed for the unevenly sampled datasets. It tests a run of trial
    periods. A light curve is first folded in a trial period and the
    variance $V_{\rm m}^2$ is computed first for each phase bin and
    then over a whole period. For a false trial, $V_{\rm m}^2$ is
    almost constant but for a true one it reaches a minimum. Instead
    of the normalized $V_{\rm m}^2$ itself, a parameter $f = (1 -
    V_{\rm m}^2) / V_{\rm m}^2$ is usually introduced to described the 
    significance of a period in a light curve \citep{kid92,liu95}. It
    is suggested that $f > 0.25 $ implies a strong period. A
    more general description and a robust test of the method was given
    by \citet{liu97}. The analysis results with Jurkevich
    $V_{\rm m}^2$ method are given in Fig.~\ref{0235j}. In the $V_{\rm
    m}^2$-plot with very high signal-to-noise ratio, there are several
    very prominent minima with broad width at the trial 
    periods $P=1.78 \, {\rm yr}$, 3.46 yr, 5.34 yr, 7.90 yr, 11.60 yr
    and 15.10 yr. The last period may not be real, because it is very
    long and the data sample covers it less than two times. Among all
    the six QPOs detected with the Lomb power spectrum method, strong
    QPOs of periods $P_1$, $P_2$, $P_3$, and $P_6$ with {\it rms}
    larger than 7 per cent are clearly detected by the Jurkevich
    $V_{\rm m}^2$ method, though the central frequencies of the
    periods shift toward to the lower end and cannot be determined
    accurately due to large noise. Weak QPOs $\nu_4$ and $\nu_5$
    with $rms \la 6\%$ are insignificant in the normalized $V_{\rm
    m}^2$-plot and are nearly drowned out by noise. In
    Fig.~\ref{0235j}, a period $P=7.90 \, {\rm yr}$ is significant but
    not present in the Lomb periodogram shown in
    Fig.~\ref{0235l8c}. One possibility is that the QPO at $P_1 =
    12.02 \, {\rm yr}$ with low quality factor {\it Q} in the Lomb
    periodogram may consist of 
    several QPOs (e.g. those periods $P= 7.90$, 11.6 yr, and 15.10 yr
    in the normalized $V_{\rm m}^2$-plot) and cannot be resolved due to
    the very low quality factor {\it Q}. However, it is more likely
    that the period $P=7.90 \, {\rm yr}$ in Fig.~\ref{0235j} does not
    represent a real QPO but just a spurious period
    at the position of two and four times of the two strong QPOs
    $P_3 = 3.62\, {\rm yr}$ and $P_6 = 1.81 \, {\rm yr}$,
    respectively. Or, the period $P = 7.90 \, {\rm yr}$ is just the
    result of incorrect identification of noise.

%----------------
%---------------

\begin{figure*}
   \plotone{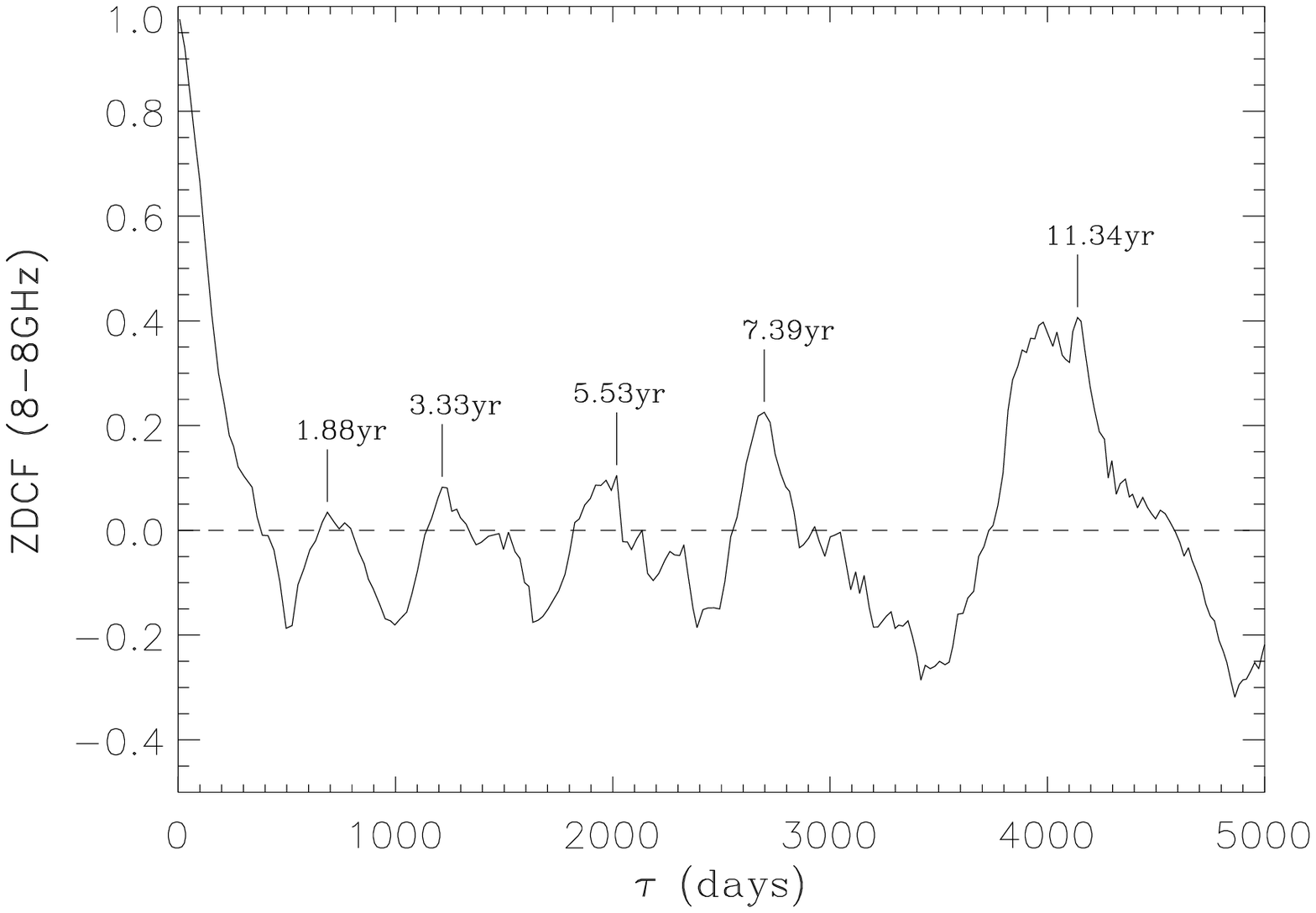}
   \caption{The periodic analysis results with ZDCF
     method. The results are consistent with that of Jurkvich $V_m^2$ 
     method and the two weakest QPOs at Lomb's periodogram are
     missing.} 
   \label{0235zd}
   \end{figure*}

    We also analyze the combined radio light curve with the
    z-transform discrete correlation function (ZDCF) technique
    \citep{ale97}, which is designed properly to treat the unevenly
    sampled datasets. Discrete self-correlation function has been
    successfully used to search for periods in the light curves of AO
    0235+164 \citep[e.g.][]{rai01} and ZDCF is basically the same as
    DCF. The self-ZDCF function of the combined radio light
    curve at 8 GHz is showed in Fig.~\ref{0235zd}.
    In self-ZDCF, a local maximum implies a period. In Fig.~\ref{0235zd}, 
    all the maxima have very broad profile and the period is
    difficult to determine accurately. Because our purpose is to
    confirm the periodic analysis results obtained with the Lomb power
    spectrum method and the Jurkevich method, we do not try to 
    compute periods from ZDCF accurately but estimate a period at 
    peak. The identified five QPOs have periods $P = 11.34 \, {\rm
    yr}$,  $7.39 \, {\rm yr}$,  $5.53 \, {\rm yr}$,  $3.33 \, {\rm
    yr}$,  $1.88 \, {\rm yr}$. The periodic analysis results with the
    ZDCF and Jurkevich methods are  
    consistent with each other within the estimate errors but not
    fully consistent with the results with the Lomb power spectrum
    method. Like the Jurkevich method, ZDCF does not detect the
    two weak periods $P_4=2.83 \, {\rm yrs}$ and $P_5=2.15 \, {\rm 
    yrs}$ obtained with Lomb's power spectrum method. It might be 
    due to the broad profiles of the strong periods $1.88\, {\rm yrs}$
    and $3.33 \, {\rm yrs}$ at ZDCF and/or to the weakness of the two
    periods, so that the two weak QPOs are absent in the ZDCF.

    The conclusions from three different periodic analysis methods 
    are that the strong QPOs with periods $P= 12.02$, 5.45, 3.62, and
    1.81 yrs are detected with unprecedentedly high singal-to-noise
    ratio in the combined 8 GHz radio light curves. Two weak QPOs
    with $P = 2.83\, {\rm yr}$ and 2.15~yr are detected by the Lomb
    power spectrum method but are not detected by the other two
    methods due to the weakness. They will be confirmed with the
    periodic investigation of radio light curves at multiple radio
    frequencies in Sec.~\ref{ssec:fredep}. 

%----
%----

%------------------------------------------------------------

\section{Harmonic relationship and dependence of QPOs on 
                 radio frequencies}
\label{resmult}

\subsection{Harmonic relationship of QPOs}

\begin{figure*}
      \epsscale{0.8}
   \plotone{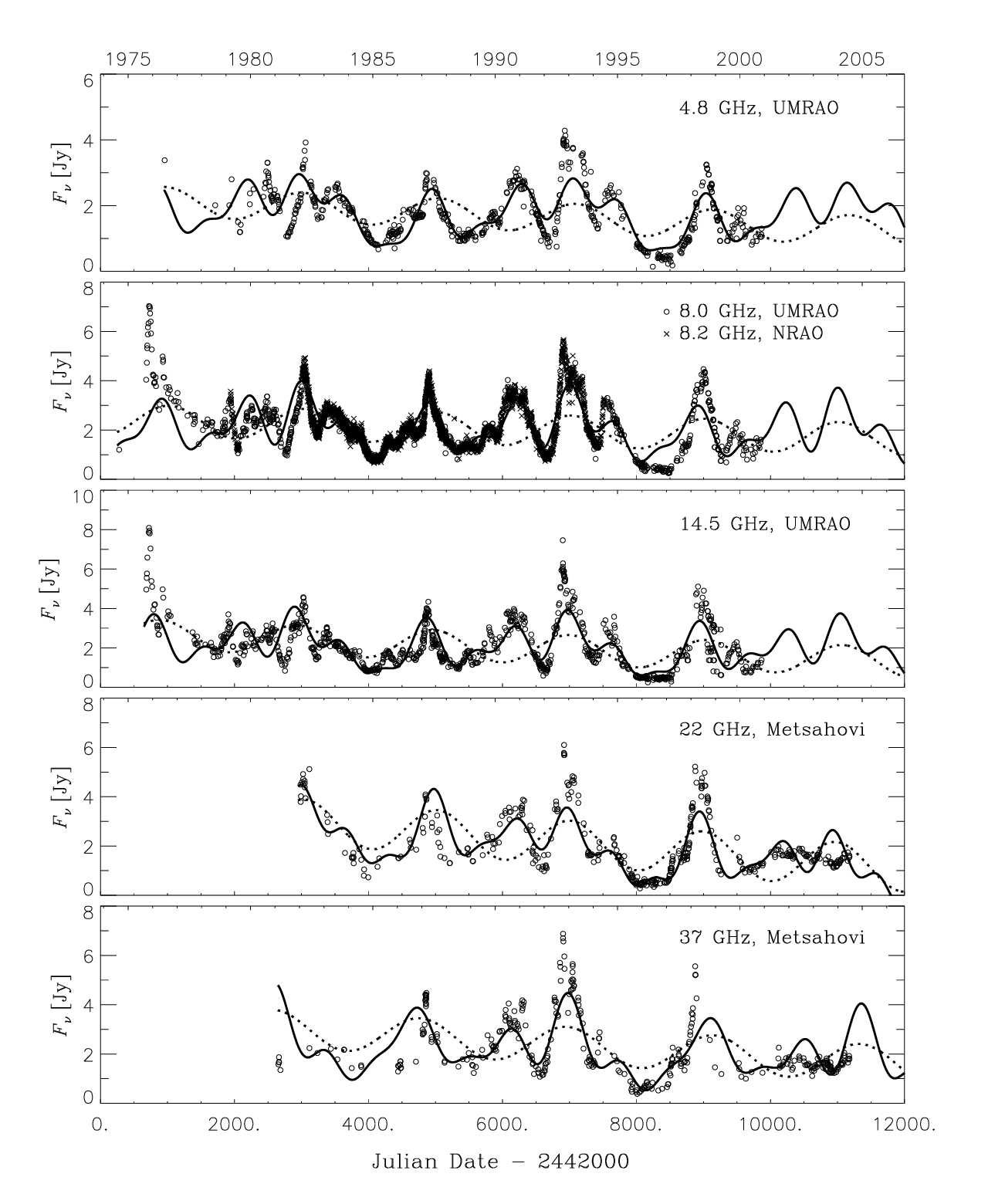}
   \caption{The radio light curves of AO 0235+164. The observation
   frequencies are indicated at the upper right corner in the
   figures. The solid line is the result fitted with six QPOs
   and the dotted line is the result fitted with the fundamental QPO
   of period $P_{\rm s}$. The results clearly showed that in addition 
   to the fundamental QPO the other five QPO harmonics significantly
   contributed to the variations of the radio light curves.}
   \label{fit8}
   \end{figure*}

    The ratio of centroid frequency and the fundamental frequency,
    $f_\nu = \nu /\nu_{\rm s}$, of six QPOs is given in Col.~8 in
    Table~\ref{idf}. We do not compute the error for the ratio as the 
    standard errors $\sigma$ of the
    frequencies given in Col.~2 are limited by the time coverage of 
    the data sample. The ratios $f_\nu$ of six
    QPOs can be given with a simple relation $f_\nu = I /
    2$ with $I=1$, 2, 3, 4, 5, and 6, or all six QPOs have
    harmonic relation $\nu_1:\nu_2:\nu_3:\nu_4:\nu_5:\nu_6 =
    1:2:3:4:5:6$. The identification is given in Col.~9 of
    Table~\ref{idf}. With the identification, we calculate the
    averaged fundamental frequency $\bar{\nu}_{\rm s}$ and the
    standard error $\bar{\sigma}_{\rm s}$ with $\bar{\nu}_{\rm s} =
    \bar{\sigma}_{\rm s} \sum (\nu_j/\sigma_j) / N $ and
    $1/\bar{\sigma}_{\rm s} = \sum (j/\sigma_j) / 2N$ with $j=2, \,
    \dots$, and 6, where N is the total number of QPOs used in the
    calculation. In the computation, we do not take into account the
    QPO with the lowest frequency $\nu_1$ (the longest period $P_1 =
    12.02 \, {\rm yr}$) due to the low quality factor $Q = 1.76$. The
    averaged frequency of the fundamental QPO is $\bar{\nu}_s =
    (0.5014 \pm 0.0434) \times 10^{-3} \, {\rm day}^{-1}$ (period $P_s
    = 5.46 \pm 0.47 \, {\rm yr}$). With the least square method, we fit 
    the combined 8 GHz radio light curve both 
    with the fundamental QPO alone ($\chi^2 = 0.7539$) and with the
    six detected QPOs ($\chi^2 = 0.3235$), and give the results 
    in Fig.~\ref{fit8}.  Fig.~\ref{fit8} shows that the fundamental
    QPO alone cannot satisfactorily reproduce the radio light curve
    while the fit with six QPOs can produce it very well until 
    about the year 2000. The results imply that the contributions
    from the harmonic QPOs are significant, and the sub-peaks of the
    outbursts are not random but have the same physical origin as that 
    of the major outbursts. The differences of the initial
    phases of six QPO harmonics relative to the fundamental 
    are listed in the last column of Table~\ref{idf}. The results
    show that the QPOs $\nu_1$, $\nu_2$, $\nu_3$, $\nu_5$, and
    $\nu_6$ almost have the same initial phases with $\Delta\phi_0
    \sim 0$ while  $\Delta\phi_0 \sim \pi$ for the 4th QPOs $\nu_4$.
    The initial phase difference $\Delta\phi_0$ of the 1st and the 5th
    QPOs are deviated from zero because the large fit
    uncertainties due to the low quality factor {\it Q} for the 1st
    harmonics or to the weakness ($rms = 4.70 \%$) of the 5th
    harmonics, respectively. The relative initial phases of QPOs are
    consistent with the suggestion of a harmonic relationship among QPOs. 

%-------------------------
%--------------------

%-----
\subsection{Frequency dependence of harmonic QPOs}
\label{ssec:fredep}

\begin{figure*}
   \plottwo{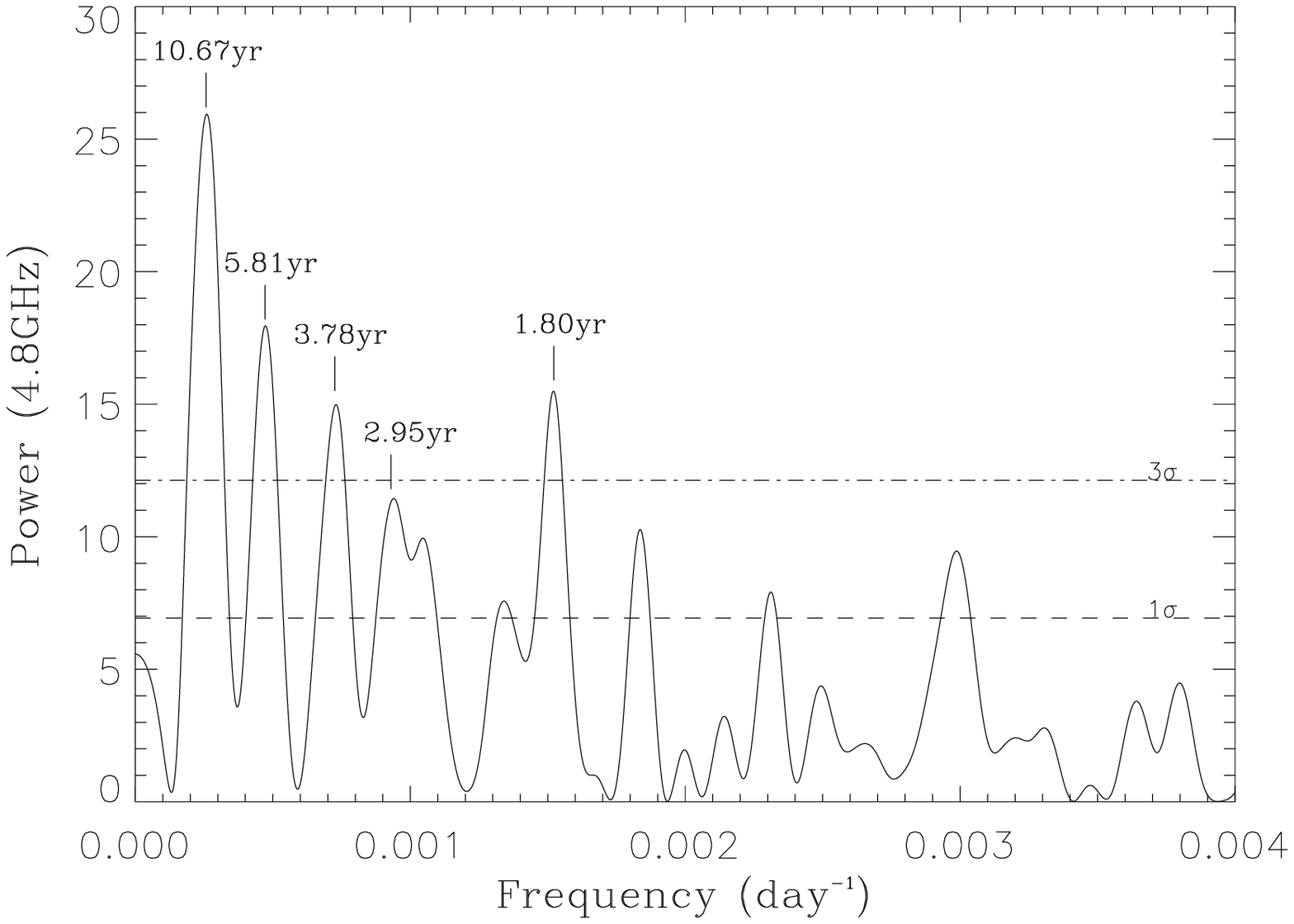}{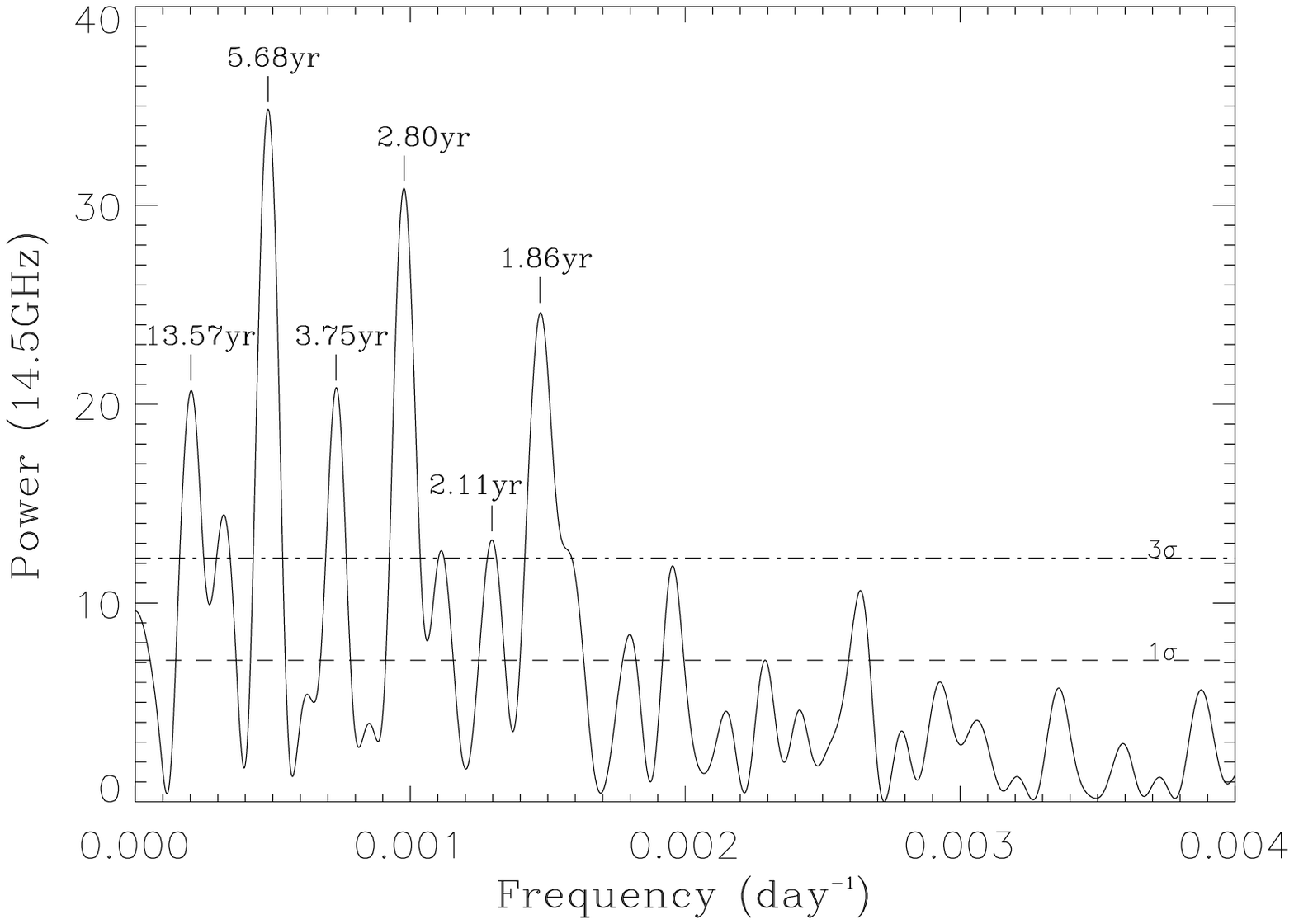}\\
   \plottwo{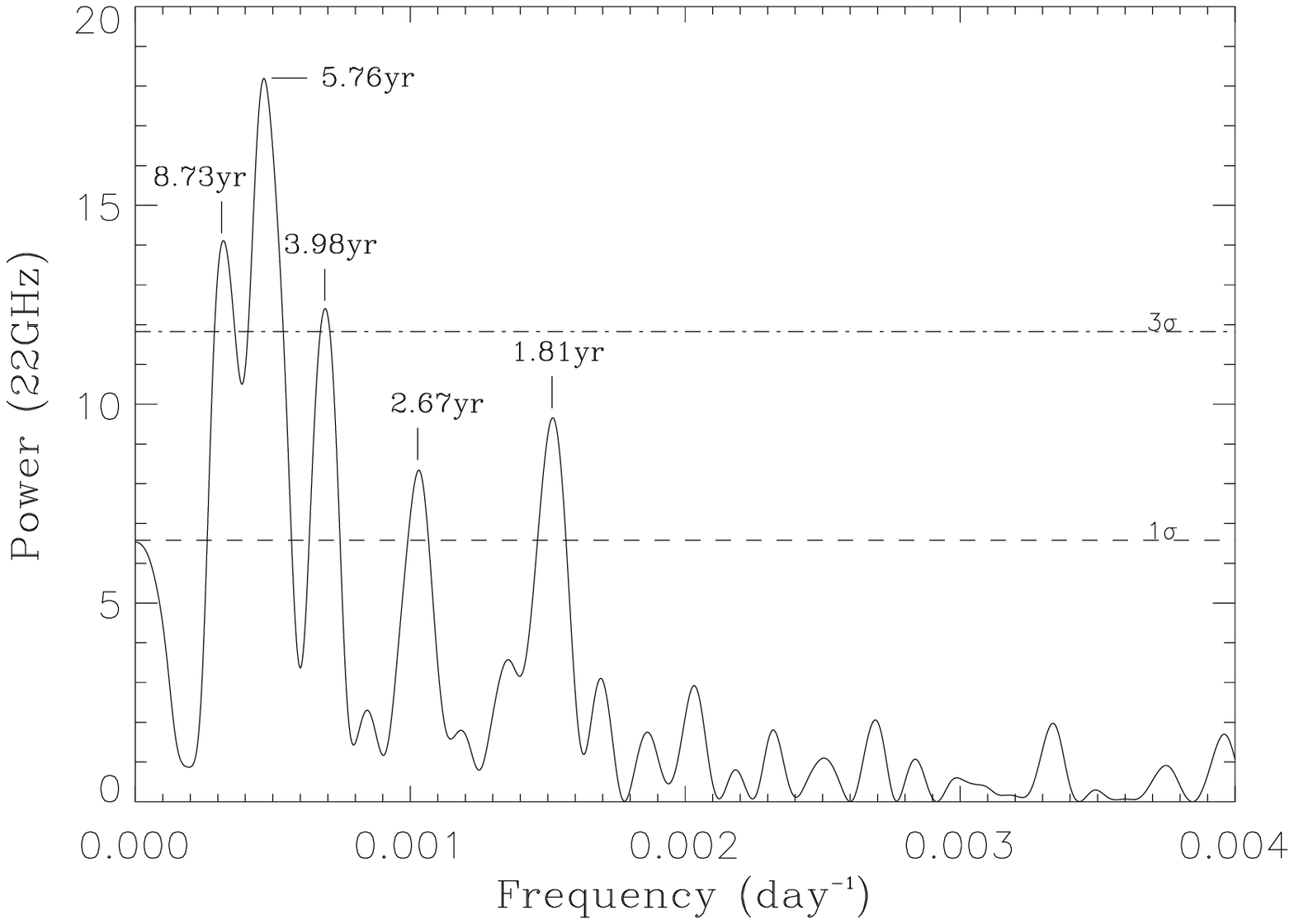}{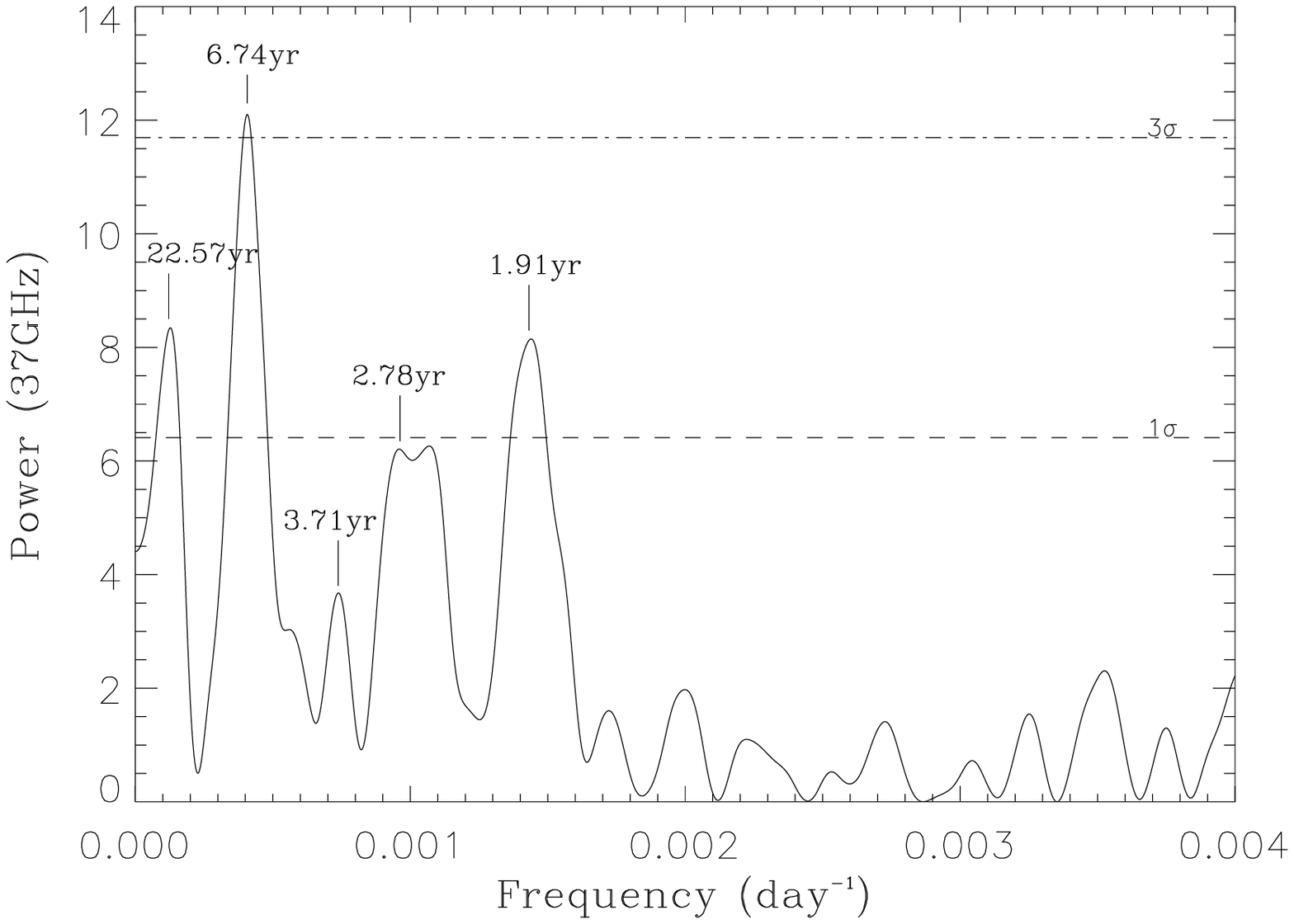}
   \caption{Lomb periodograms of the radio light curves at 4.8 GHz 
     ({\it uper left}), 14.5 GHz ({\it uper right}), 22 GHz ({\it
     lower left}), and 37 GHz ({\it lower right}). Fitted centroid
     periods of QPOs with Gaussian function are indicated. The dashed
     and dash-dotted lines are the $1 \sigma$ and $3 \sigma$
     significance levels, respectively. The radio data at 4.8 GHz and
     14.5 are from UMRAO and at 22 GHz and 37 GHz from Mets\"ahovi
     Observatory.} 
   \label{0235loht}
   \end{figure*}

    In this section, we present QPO analyses with the radio light curves
    at 4.8 GHz, 14.5 GHz, 22 GHz, and 37 GHz and investigate the
    dependence of the harmonic QPOs on radio frequency. Since
    our purpose is to investigate the QPOs and the frequency
    dependence, we give only the analysis results with the Lomb power
    spectrum method. The radio data at 4.8 GHz 
    and 14.5 GHz from UMRAO and at 22 GHz and 37 GHz from Mets\"ahovi
    Observatory are plotted in Fig.~\ref{fit8}. The periodic analysis
    results are given in Fig.~\ref{0235loht}.

%--------------
%-----------------------

    Fig.~\ref{0235loht} shows that the Lomb periodograms of radio
    light curves at the four radio frequencies are very noisy and
    the signal-to-noises of the Lomb periodogram are significantly
    lower than that of the combined radio light curve at 8 GHz. Among
    six QPOs obtained at 8 GHz, the five strongest QPOs are
    detected in the radio light curves at all the radio frequencies,
    while the weakest QPO of period $P_5 = 2.15 \, {\rm yr}$ and 
    $rms = 4.70$~per cent at 8 GHz is significantly detected only at
    14.5 GHz, at which frequency the radio light curve is 
    relatively well 
    sampled. At 4.8 GHz, 22 GHz and 37 GHz, the weakest QPO is absent 
    from the Lomb power spectra probably due to poor sampling and to 
    the missing of the information on the minor events. Like at 8 GHz,
    the peak at a period with $P \sim 7 \, {\rm yr}$ obtained both with 
    the Jurkevich $\rm V_m^2$ and the ZDCF methods, is not present in
    the Lomb periodograms at all the radio frequencies in
    Fig.~\ref{0235loht}. Table~\ref{idf} 
    lists the fitted centroid frequency with Gaussian function,
    fitting goodness $\chi^2$, relative strength {\it rms}, quality
    factor {\it Q}, the ratio of the frequencies of harmonics and the
    fundamental QPOs,  and the averaged fundamental period $\bar{P}_{\rm
    s}$. Again, the QPOs clearly have a harmonic relationship and the
    identification of the harmonic integer is given in
    Table~\ref{idf}. Like what we did for the combined 8~GHz  
    light curve, with the least square method we also fit 
    the radio light curves at 4.8 GHz, 14.5 GHz, 22 GHz and 37 GHz
    both with the fundamental QPO $\bar{\nu}_{\rm s}$ alone and the
    six QPOs 
    together, respectively. The fitted light curves are given in
    Fig.~\ref{fit8} and the differences of the initial phases of
    harmonic and the fundamental QPOs are listed in Col.~10 of
    Table~\ref{idf}. From the fitting results in Table~\ref{idf},
    the differences of initial phases of the six QPOs at the four
    radio frequencies are consistent with the results obtained at 8
    GHz and five QPOs $\nu_1$, $\nu_2$, $\nu_3$, $\nu_5$, and
    $\nu_6$ have almost the same initial phases but about $\pi$
    difference from that of the 4th one. The results strongly
    support the identification of six QPOs and the harmonic 
    relationship. 

%---------------
\begin{figure*}
\plotone{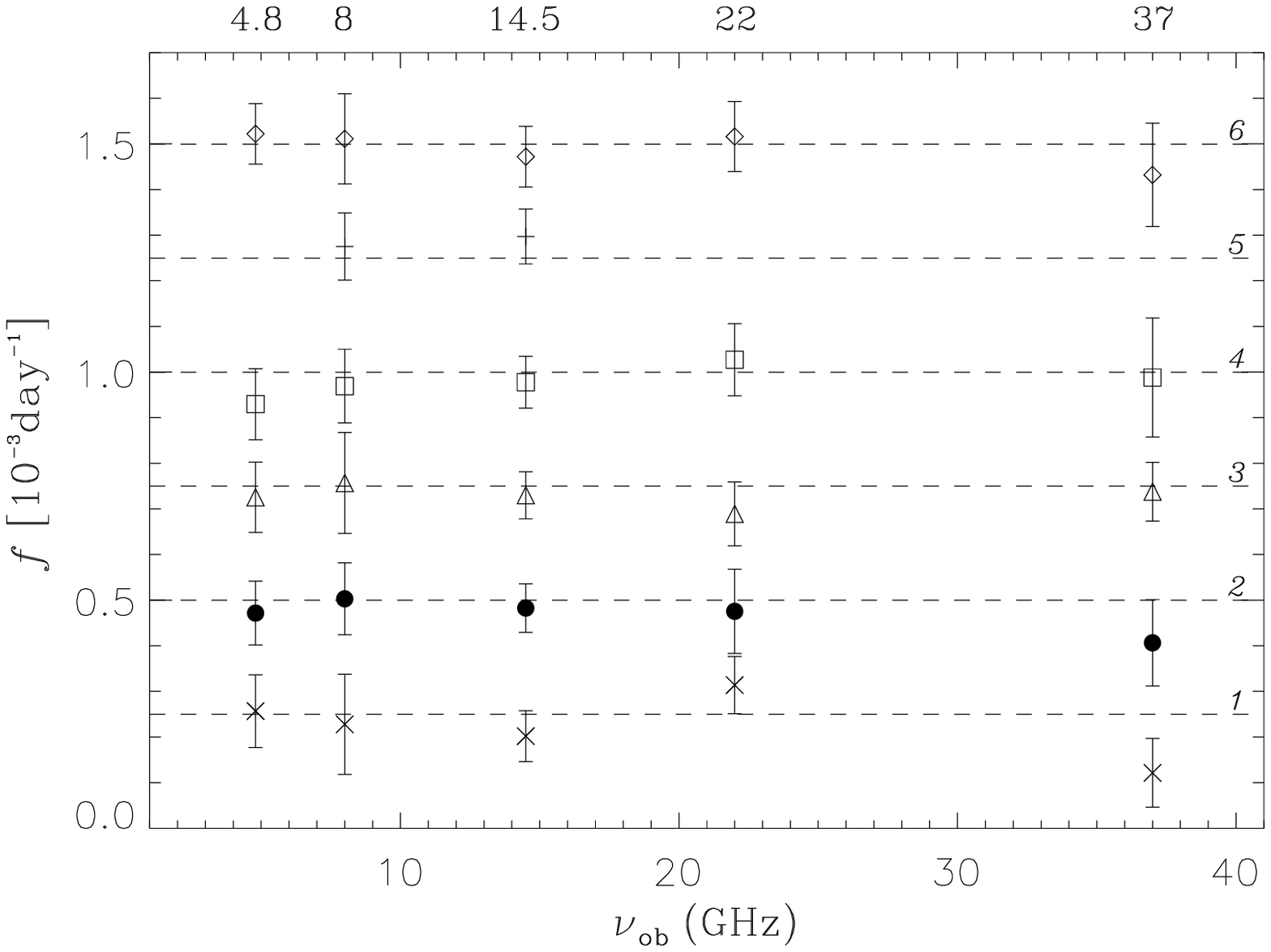}
   \caption{Dependence of QPO frequencies on radio frequencies. The
   horizontal dashed lines with small integer number $N=1$, 2, 3, 4,
   5, and 6 indicates the expected ratio of frequencies of QPOs and
   the basic QPO with lowest frequency (calculated basing the
   fundamental QPO frequency $\nu_{\rm s}$ at 8 GHz). The centroid
   frequencies and the harmonic relationship of QPOs 
   are independent of the observational radio frequencies.}  
   \label{qpofreq}
\end{figure*}
%--------------

\begin{figure*}
   \plotone{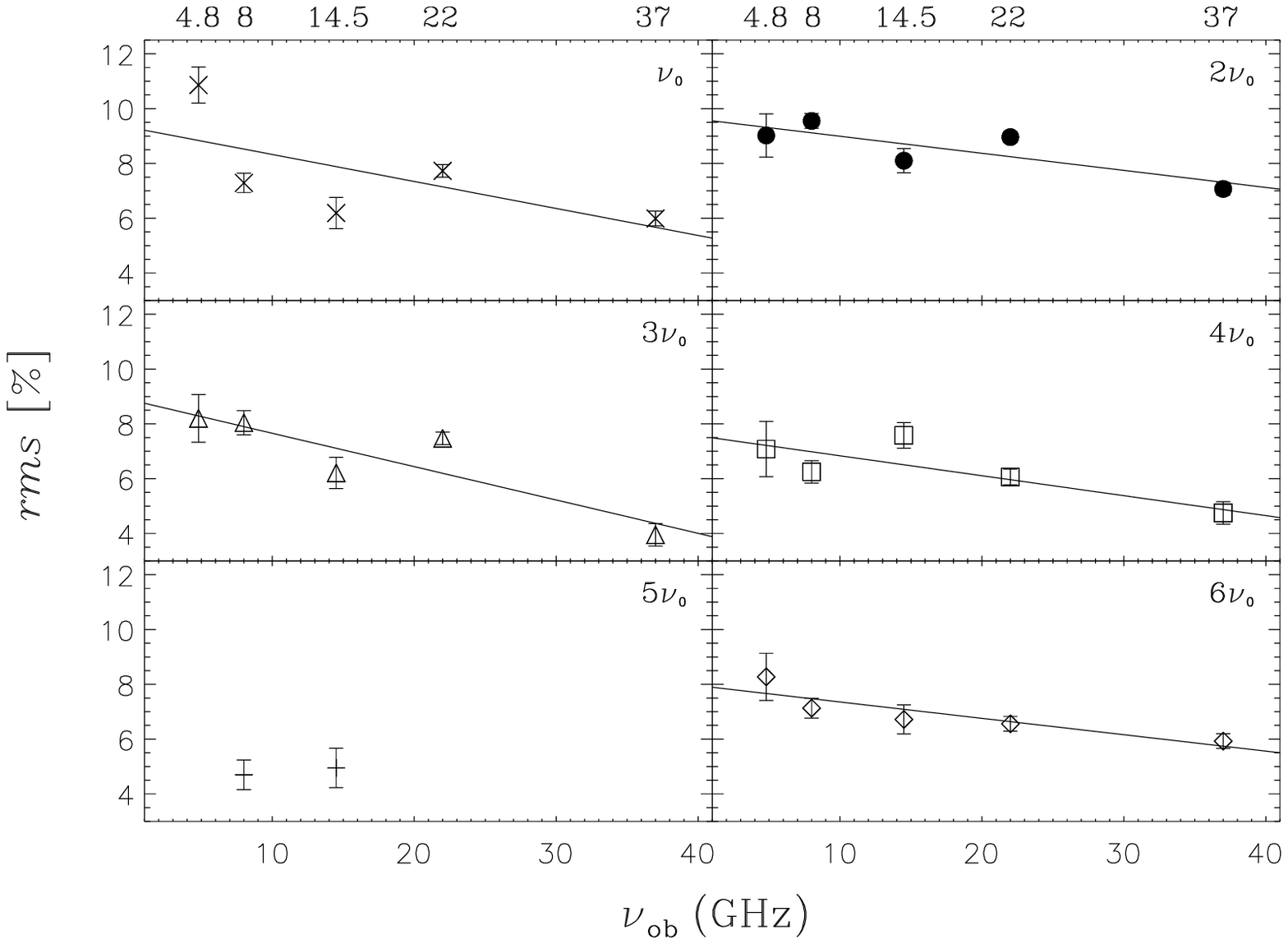}
   \plotone{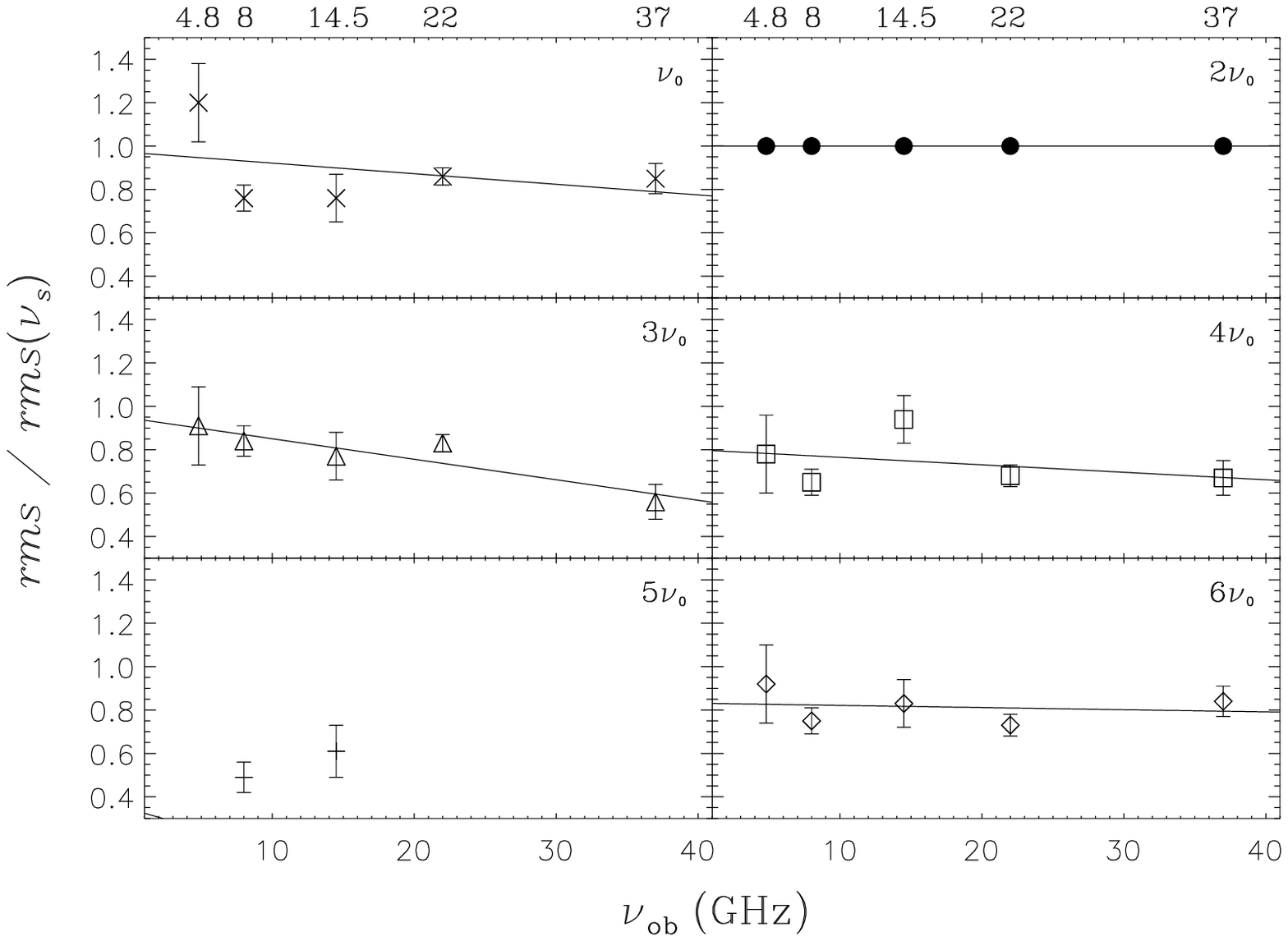}
   \caption{The amplitude {\it rms} (in percentage) of QPO harmonics
   as a function of the radio frequencies ({\it upper
   panel}) and the amplitude {\it rms} relative to that of the
   fundamental QPO (the 2nd harmonics) {\it vs} radio frequencies
   ({\it lower panel}). The solid line is the fitted results with
   least square method.} 
   \label{freqrms}
\end{figure*}

    The dependence of QPO centroid frequency on the observational radio
    frequencies is given in Fig.~\ref{qpofreq}. Fig.~\ref{qpofreq}
    shows that QPO frequencies are independent of the  radio
    frequencies and the ratios of the frequencies of harmonic QPOs at
    the four radio wave-bands and the fundamental one with $\nu_{\rm s}
    = 0.5029 \times 10^{-3} \, {\rm day}^{-1}$ at 8 GHz are also 
    independent of the radio frequencies, implying that the QPO
    centroid frequencies and the harmonic relationship are not
    determined by the different radio emission regions in 
    relativistic jet. However, the independence does not mean that one
    could detect all the harmonic components of QPOs at all
    wave-bands. To answer this question, we investigate the
    dependence of relative amplitude {\it rms} on radio
    frequencies. Fig.~\ref{freqrms} shows the results that the
    relative amplitude {\it rms} of all QPOs is a function of radio
    frequency and the 
    third QPO has the strongest dependence. However, the conclusion is 
    very sensitive to the periodic analysis results of radio light
    curves at 37 GHz, which is poorly sampled. Therefore, the
    decrease of QPO amplitudes may be due to the loss of information
    on high frequency QPOs in the poorly-sampled radio light
    curves. To eliminate the effect of the incomplete coverage of
    outburst structures on the analysis results of QPO harmonic
    components, we calculate the ratio of {\it rms} of the harmonics
    and $rms(\nu_{\rm s})$ of the fundamental QPO. The results are
    given in Fig.~\ref{freqrms}. In Fig.~\ref{freqrms}, $rms / rms ({\rm
    \nu_s})$ for the 1st, 4th and 6th components of the harmonic QPOs
    is nearly independent of frequency while for the 3rd harmonics the
    ratio decreases with radio frequency. Again, the conclusion is
    sensitive to the periodic analysis results of the 37 GHz light
    curve. If the analysis results of QPOs at 37 GHz are not
    considered, the harmonics of QPOs are nearly independent of 
    radio frequencies. The results suggest that if one detects the
    fundamental QPO of period $P_{\rm s} = 5.46 \, {\rm yr}$ in a
    well-sampled light curve at high frequency, e.g. in the optical
    wave-bands, one would 
    simultaneously detect the other QPOs with periods $P_1 \simeq 
    10.9 \, {\rm yr}$, $P_4 \simeq  2.8 \, {\rm yr}$, and $P_6 
    \simeq 1.8 \, {\rm yr}$. But, the QPO of period $P_3 = 3.6 \, {\rm 
    yr}$ may not be observable in optical wave-band. Although the
    conclusion needs to be tested with much more observations at 37
    GHz, our results suggest that the properties of QPOs would
    be independent of radio frequency and not be determined by the
    physical process in the radio emission regions of jet.

%-------------------
\section{Central black hole mass and thick disk oscillations}
\label{discuss}

    \citet{rai01} reported the detection of a period $P = 5.7 \pm 0.5
    \, {\rm yrs}$ in 
    the optical and radio light curves of AO 0235+164. 
    \citet{osto04} interpreted it with a helical
    jet model, in which the quasi-periodic radio-optical light curve
    and the spectral energy distributions (SEDs) are caused by the
    orientation variation of a helical inhomogeneous, non-thermally
    emitting jet probably due to the orbital motion of the primary
    black hole in a binary system. In the helical jet model, the
    minor burst events are taken as non-periodic flux fluctuations
    due to the random plasma instabilities. However, our periodic
    analyses show that the major outbursts and the minor burst events
    are the products of combinations of six harmonic QPOs, which
    suggest a quasi-periodic radio emission from relativistic
    jets and a physical origin of harmonic QPOs independent of the
    radio emission regions in the jet. The multiplicity, the harmonic
    relationship, the zero differences of initial phases, and the
    independence of relative {\it rms} on radio frequency imply that
    the harmonic QPOs in the radio light curves of AO 0235+164 are
    most likely due to the quasi-periodic injection of plasma from
    accretion disk into the relativistic jet. Because of the jet-disk
    coupling \citep{fend04}, such a quasi-periodic plasma injection
    implies a quasi-periodic oscillation of accretion around SMBH.

    Both thin and thick accretion disks can oscillate
    quasi-periodically, but only the oscillation of a thick disk can
    be global and trigger the quasi-periodic accretion with harmonic
    frequencies. \citet{rezzolla03} and \citet{zanotti03} 
    found that the accretion rate of a tori or thick disk of finite
    radial extent is quasi-periodic due to the global p-mode
    oscillation, which could be excited by a global perturbation 
    \citep{rezzolla03b,zanotti03} or a local periodic strong agent 
    \citep{rubio05a,rubio05b}. Such a torus or thick disk of finite
    radial extent could be a optically thin and geometrically thick
    radiatively inefficient accretion flow (RIAF) or advection
    dominated accretion flow (ADAF) between a central supermassive
    black hole and an outer truncated geometrically thin disk.  It is
    clear now that such a configuration should be present in a black
    hole accretion system if the dimensionless accretion rate $\dot{m}
    \equiv \dot{M} /\dot{M}_{\rm Edd}$ is lower than a critical value
    $\dot{m}_{\rm cr} \simeq 0.02$ \citep[e.g.][]{meyer94,narayan95,
      narayan02,esin97,meyer00, meyer05}, where $\dot{M}_{\rm Edd} =
    L_{\rm Edd}/0.1 c^2$ is the Eddington accretion rate relative to
    the Eddington luminosity $L_{\rm Edd} \simeq 1.26\times 10^{46}
    \, {\rm erg\; s^{-1}} \left(M / 10^8 M_\odot\right)$. Because the
    p-mode oscillation frequency of thick accretion disk is determined
    by the disk radial extent defined with the dimensionless accretion
    rate, we now estimate the mass $M_{\rm BH}$ of central black hole
    and then the dimensionless accretion rate $\dot{m}$ of AO
    0235+164.  

%--------------------------------------------
\subsection{Black hole mass and accretion rate}
\label{sec:bhmass}

    As there is no valid estimation of the black hole mass of AO
    0235+164 in literature, we estimate its black hole mass
    and the accretion rate in this section,  before we start the
    investigations of the physical origin of QPOs. Because 
    of the high redshift ($z= 0.94$) and the very bright central
    nucleus, it is impossible to resolve the host galaxy of AO
    0235+164 even with the
    Hubble Space Telescope and to infer the black hole mass
    with the tight relation of black hole mass and the properties of
    galaxy bulge (e.g. stellar velocity dispersion $\sigma$). However, we 
    can estimate the black hole mass with some recently suggested relations 
    of the emission line properties and central black hole mass in
    AGNs.

    The reverberation mapping studies reveal an empirical relation
    of broad emission line region (BLR) size and the optical continuum
    luminosity at rest-frame 5100$\AA$ \citep{kaspi05} 
\begin{equation}
  {R_{\rm BLR} \over 10 \, {\rm lt-days}} = (2.23 \pm 0.21) \left[
      {\lambda L_\lambda (5100\AA) \over 10^{44} \, {\rm ergs~s^{-1}}}
      \right]^{0.69\pm0.05} .
\label{eqblrlum}
\end{equation}
    Together with the FWHM of H$_\beta$ emission
    line, which is a measurement of the characteristic velocity of
    broad emission line clouds, one can estimate the central black
    hole mass of an AGN with the relation \citep{peterson04}
\begin{equation}
  M_{\rm BH}=1.936 \times 10^5 M_\odot \left( {R_{\rm BLR} \over {\rm
      lt-days}}\right) \left[ {\rm FWHM(H_\beta) \over 10^3 \,
      km~s^{-1}}\right]^2.  
  \label{eqmsl}
\end{equation}

    The environment of AO 0235+164 is rather complex due to the
    presence of several foreground intervening galaxies within a few
    arc-seconds. Great efforts have been done to study the 
    emission line features of the object with spectroscopic
    observations. Using the 3m Shane telescope at Lick Observatory,
    \citet{cohen87} identified the emission lines of MgII, [NeV] and
    [OII] at the source red shift. With the 4m William Herschel
    Telescope, \citet{nilsson96} detected not only the [NeV] and [OII]
    lines but also prominent hydrogen emission lines H$_\delta$ and
    H$_\gamma$ at the same redshift. \citet{nilsson96} obtained the
    FWHM and flux (in the observer's frame) of H$_\delta$($\lambda
    4102$) and H$_\gamma$($\lambda 4340$) emission lines to be $\rm
    3600\pm400 \, km~s^{-1}$, $39\times 10^{-17} \, {\rm erg
    s^{-1}cm^{-2}}$ and $3400\pm400 {\rm km~s^{-1}}$,
    $97\times10^{-17} \, {\rm erg~s^{-1}cm^{-2}}$, respectively. The
    equivalent widths of the H$_\delta$ and H$_\gamma$ lines are
    2.6$\AA$ and 7.2$\AA$ (all the quantities are given in rest frame
    except noted). With the assumption of a power-law optical
    continuum ($F_\nu \propto \nu^{-\alpha}$), we can estimate the flux
    at 5100$\AA$ from the flux in the bluer band. However, deriving an
    accurate value of spectral index $\alpha$ is not a simple task for
    AO 0235+164 because of its strong variation \citep{smith87}.
    A visual inspection of Fig. 1 of \citet{cohen87} reveals that
    at 4500$\AA$ (observer's frame) the continuum flux density is
    about 0.5$\times 10^{-16} {\rm erg~s^{-1}cm^{-2}}\AA^{-1}$, while
    at 7500$\AA$ it is about 1.1$\times 10^{-16} {\rm
    erg~s^{-1}cm^{-2}}\AA^{-1}$. From these flux values we can derive
    $\alpha$=3.54, which is in fair agreement with the values of 
    $\alpha\sim 3.6$ corresponding to a faint state of the source
    reported in Fig. 7 of \citet{rai01}. However, when one takes into
    account the absorption by the intervening system at z=0.524, this
    value nearly halves. This can be inferred by the intrinsic
    (dereddened) $<B-R>=1.72$ derived by \citet{rai05} which, when
    considering the amount of extra absorption tabulated in their
    Table 5, gives $<\alpha>=1.75$. We notice that this $\alpha$ value
    means a relatively flat shape in the optical band of the flux
    density versus wavelength plot ($F_\lambda \propto
    \lambda^{\alpha-2}$), which is also consistent with the
    dereddenned spectrum shown in Fig. 7 of \citet{junkk04}. 
    Adopting $\alpha$=1.75, from the observed flux at 7500$\AA$
    (rest-frame 3866$\AA$) in Fig. 1 of \citet{cohen87} we calculate
    the  flux density at rest-frame 5100$\AA$ as 7.5$\times 10^{-16}
    {\rm erg~s^{-1}cm^{-2}}\AA^{-1}$, while from the continuum flux at
    4102$\AA$ estimated from the H$_\delta$ line flux and equivalent
    width in Nilsson et al. (1996) we calculate the  flux density at
    rest-frame 5100$\AA$ as 5.3$\times 10^{-16} {\rm
    erg~s^{-1}cm^{-2}}\AA^{-1}$. From the average value of these two
    estimates, we obtain the monochromatic luminosity at 5100$\AA$
    with $\lambda L_\lambda (5100\AA)= 7.48 \times 10^{45} \, {\rm
    erg~s^{-1}}$. From
    Eq.~(\ref{eqblrlum}), we estimate the BLR size as $R_{\rm 
    BLR}= 438\pm103  \, {\rm lt-days}$. Because of no
    detection of H$_\beta$ for AO 0235+164 in literature, we try to 
    estimate its FWHM from the observations of other Hydrogen Balmer 
    lines. As the observed H$_\gamma$ line is probably contaminated by
    [OIII]($\lambda 4363$) line, we make the estimation with H$_\delta$
    line data. From the observations of 50 quasars in LBQS given by
    \citet{forster01}, we derive an empirical relation between the 
    FWHM values of H$_\beta$ and H$_\delta$ lines with the OLS
    bisector method \citep{isobe90} 
\begin{equation}
  \rm FWHM(H_\beta) = (415.6 \pm 348.8) + (1.07 \pm 0.15)
  FWHM(H_\delta) .
  \label{eqfwhmbd}
\end{equation}
    From the observations of the FWHM of H$_\delta$ line, we have $\rm
    FWHM(H_\beta) \simeq 4267\pm772 \, km~s^{-1}$ for AO 0235+164. From
    Eq.~(\ref{eqmsl}), we estimate the central black hole mass of AO
    0235+164 to be $M_{\rm BH} \simeq (1.54\pm0.67) \times 10^9M_\odot$.
    However, AO 0235+164 is a blazar-type object and its optical
    continuum luminosity is generally dominated by the beamed
    synchrotron emission from a relativistic jet rather by the
    emission from accretion disk. The southern intervening galaxy with
    z=0.524 may also partly contribute to the optical continuum
    luminosity \citep{nilsson96}. Therefore, the estimated black hole
    mass with the optical continuum luminosity is overestimated
    and can be taken only as an upper limit. 

    To avoid such an overestimation of black hole mass, \citet{wu04}  
    argued that the emission line luminosity is probably a better tracer 
    of the ionizing continuum luminosity of blazars and proposed a 
    new empirical relation between
    the BLR size and the $H_\beta$ emission line luminosity  
\begin{equation}
  \lg R_{\rm BLR}({\rm lt-days}) = (1.381 \pm 0.080) + (0.684 \pm
  0.106) \lg (L_{H_\beta}/10^{42} \, {\rm ergs~s^{-1}}) .
\label{eqblrline}
\end{equation}
    Because no detection of H$_\beta$ emission line is reported for AO
    0235+164 in literature,  we estimate the luminosity from the
    observations of H$_\delta$ 
    and H$_\gamma$ line fluxes. \citet{zheng88} studied the variations
    of Balmer decrements of several quasars. From his table 2 and
    table 3, we get an averaged flux ratios 5 and 2.4 of H$_\beta$ to
    H$_\delta$ and H$_\beta$ to H$_\gamma$, respectively. 
    From the observations of emission lines by \citet{nilsson96}, we
    obtain the rest-frame H$_\beta$ luminosity for AO 0235+164 with
    $L_{H_\beta} \simeq 1.68 \times 10^{43} \, {\rm ergs~s^{-1}}$ from 
    the observed flux of H$\delta$ emission line. The estimated
    H$_\beta$ 
    luminosity from the observed H$_\gamma$ flux is about 15 per cent
    higher, which might be due to the contamination of [OIII]($\lambda
    4363$) line. From Eq.~(\ref{eqblrline}) and the H$_\beta$
    luminosity estimated from H$_\delta$ emission line flux, we
    calculate the BLR size as $R_{\rm BLR}= 166\pm64 \, {\rm lt-days}$,
    which is much smaller than that estimated from the optical
    continuum luminosity. With the estimated BLR size and 
    H$\beta$ FWHM value, from
    Eq.~(\ref{eqmsl}) we estimate the black hole mass of AO
    0235+164 of $M_{\rm BH}= (5.85\pm3.09) \times 10^8 M_\odot$, which is
    much smaller than that obtained with the optical
    continuum luminosity method.
 
    \citet{mclure04} suggested to estimate the black hole mass of AGNs
    from the UV continuum luminosity and the FWHM of
    MgII(2788\AA)  emission line with an empirical relation (with
    cosmological parameter $H_0=70\, {\rm km~s^{-1}~Mpc^{-1}}$,
    $\Omega_{\rm m}=0.3$ and  $\Omega_\Lambda=0.7$; the distance of
AO 0235+164 is 6.1Gpc in this case)
\begin{equation}
  M_{\rm BH}=3.2 M_\odot \left({\lambda L_{3000 \AA} \over
    10^{44} {\rm ers\; s^{-1}}}\right)^{0.62} \left[{\rm FWHM(MgII)
    \over km~s^{-1}}\right]^2 .
\label{eqmmgii} 
\end{equation}
    For AO 0235+164, the MgII emission line is observed to have the
    FWHM of $3100 \, \rm km~s^{-1}$ and the flux of $1.24\times
    10^{-15}\, {\rm 
    ergs~cm^{-2}~s^{-1}}$ \citep{cohen87}. The equivalent width and
    the continuum flux are $15.7\AA$ and  $0.8\times 10^{-16}\, {\rm 
    ergs~cm^{-2}~s^{-1}~\AA^{-1}}$, respectively. Using the
    observations of the MgII emission line and the continuum, we
    derive the rest-frame continuum luminosity at 3000$\AA$ of $\rm
    \lambda L_{3000\AA} = 7.37 \times 10^{45} \, {\rm erg\; s^{-1}}$
    (using the same cosmological parameters as in \citet{mclure04}). 
    From Eq.~(\ref{eqmmgii}), we obtain a black hole mass $M_{\rm
    BH}=4.42 \times 10^8 {\rm M}_\odot$, which is approximately the
    same as the estimation obtained with the luminosity of H$\beta$
    emission line. However, the same as the optical continuum
    luminosity, the UV continuum luminosity of AO235+164
    is also contributed by the jet emission and probably by the
    intervening galaxy. On the other hand, a recent
    study of \citet{junkk04} indicated that the UV continuum of
    AO 0235+16 is heavily absorbed by the Galaxy and the intervening
    materials at z=0.524, which implies that the de-reddened UV
    continuum is much larger than the observed one (see their Fig. 7).
    Therefore, the UV continuum luminosity of AO 0235+164 suffers
    large uncertainties and the estimated black hole mass derived with
    it may be unreliable and will be considered only as a reference. 

    It has been found that there is a tight correlation between the
    black hole mass and the bulge velocity dispersion for nearby
    galaxies \citep{gebhardt00,merritt01,tremaine02}, which can be
    expressed as \citep{tremaine02} 
\begin{equation}
  Log (M_{BH}/M_\odot)=8.13\pm0.06 + (4.02\pm0.32)~Log (\sigma/ 200 \, {\rm
    km~s^{-1}}) .
  \label{eqmsig}
\end{equation}
    AGNs seem to follow the same correlation as nearby galaxies
    \citep{ferrarese01,onken04}. On the other hand,
    there is also a tight correlation between the central velocity
    dispersion of galactic bulge and the FWHM of [OIII](5007$\AA$)
    emission line in AGNs with
    $\sigma \approx FWHM([OIII])/ 2.35$ \citep{nelson95}. This suggests
    that black hole mass could be roughly estimated from the observed
    FWHM values of narrow forbidden lines in AGNs. For AO 0235+164,
    both \citet{cohen87} and \citet{nilsson96} have 
    detected narrow forbidden lines at redshift of  $z=0.94$. 
    \citet{cohen87} estimated the FWHM values of [Ne V](3426$\AA$) and
    [OII](3727$\AA$) as 600 $km~s^{-1}$ and 200$km~s^{-1}$,
    respectively. \citet{nilsson96} obtained the FWHM values of [Ne
    V](3346$\AA$), [Ne V](3426$\AA$) and [OII] (3727$\AA$) as
    700$\pm$100$km~s^{-1}$,  400$\pm$100 $km~s^{-1}$ and
    600$\pm$100$km~s^{-1}$, respectively. From the emission line data
    of quasars in LBQS \citep{forster01}, we know that the average
    FWHM value of [OIII](5007$\AA$) is comparable with those of 
    [OII] (3727$\AA$) and other narrow forbidden lines. Therefore we
    roughly take 600$km~s^{-1}$ as the FWHM value of
    [OIII](5007$\AA$).  Using $\sigma \approx FWHM([OIII])/2.35$, we 
    derive the value of a central velocity dispersion $\sigma = 255\,
    {\rm km~s^{-1}}$. From Eq.~(\ref{eqmsig}), we estimate the black
    hole mass of AO 0235+164 as $M_{BH} \approx (3.58\pm2.65) \times 10^8
    M_\odot$, which is consistent with the value obtained with
    the H$\beta$-luminosity. 
   
    We have adopted four different kinds of methods to derive the
    SMBH mass of AO 0235+16. Except the optical continuum luminosity
    method which apprently overestimates the SMBH mass,
    the other three methods give the SMBH mass in a range from
    $3 \times10^8 M_\odot$ to $6\times10^8M_\odot$, with typical
    uncertainties of  a factor of a few. In the following
    discussions, we take the averaged mass $M_{\rm BH} = (4.72\pm2.04)
    \times 10^8 \, {\rm M}_\odot$ for the central black hole of AO
    0235+164. However, we must keep in mind that the value is derived 
    either with the indirect estimated H$\beta$ and [OIII](5007$\AA$)
    emission line properties or with the UV continuum luminosity which
    suffers from serious contaminations from the jet, southern
    intervening galaxy and extra-absorptions. In order to obtain more
    reliable black hole mass, further studies are absolutely needed to
    accurately determine the emission line and continuum properties 
    of AO 0235+164.

   \citet{wang04} calculated the broad line luminosity $L_{\rm BLR}
   \simeq 4.1 \times 10^{43} \, {\rm ergs~s^{-1}}$ for AO 0235+164
   (adjusted to fit the cosmological parameters used here). If we
   adopted a typical covering factor $C \simeq 0.1$ for the broad-emission
   line region \citep{netzer90}, the bolometric or the ionization
   luminosity is $L_{\rm bol} = L_{\rm ion} \simeq L_{\rm BLR}/C
   \simeq 4.1 \times 10^{44} \, {\rm 
   ergs~s^{-1}}$. From the estimated bolometric luminosity and black
   hole mass, we obtain the dimensionless accretion rate
   $\dot{m} \simeq  7\times 10^{-3} (\eta/0.1)^{-1}$ of AO 0235+164,
   where $\eta$ is the conversion efficiency of matter to energy.

%------
\subsection{Thick accretion disk p-mode oscillations and
   physical origin of multiple QPOs} 
\label{sec:diskosc}

   The estimated dimensionless accretion rate of AO 0235+164 is
   about two times smaller than the critical accretion rate $\dot{m}_{\rm
   cr} \simeq 0.02$ in accretion disk evaporation model 
    \citep{meyer00,liubf02} or much less than that of  
   $\dot{m}_{\rm cr} \simeq 0.05$ in strong ADAF principle 
    \citep{narayan95,abram95}, implying
   that the central region of the accretion disk in AO 0235+164 
   is in ADAF (or RIAF) and the outer region of the accretion 
   disk is a truncated standard thin disk. There is no consensus in 
   literature about how to determine the transition
   radius $R_{\rm tr}$ of inner thick ADAF and an outer thin standard
   disk. Three kinds of methods based on different physical processes
   have been proposed: disk evaporation
   \citep{meyer94,liu95b,meyer00,liubf02,
     rozanska00}, the absence of a strong ADAF solution
   \citep{narayan95,abram95}, and disk instability of inner 
   radiation pressure dominated region with energy transfer outwards
   \citep{honma96,kato98,lu04}. The three methods generally give very
   different estimation of the transition radius. In the 
   evaporation model, the transition radius
   at the critical accretion rate $\dot{m}_{\rm cr}$ is
   $R_{\rm tr} \simeq 355 r_{\rm G}$ \citep{meyer00,liubf02}, while in 
   the strong ADAF scenario, the transition radius at the critical
   accretion rate $\dot{m}_{\rm cr} \simeq 0.05$ is about $R_{\rm tr}
   \simeq 560 r_{\rm G}$  \citep{narayan95}. The constraints 
   from the existence of the broad line region in AGNs favored the 
   relation of transition radius and accretion rate given by the
   strong ADAF principle \citep{czerny04}. 

   In the strong ADAF scenario with advection fraction of energy
   $f=0.9$, the relation of the transition radius and accretion rate  
   for $\dot{m} < \dot{m}_{\rm cr}$ is $R_{\rm tr} \simeq 0.074
   \left({\alpha / 0.13}\right)^4 \dot{m}^{-2} r_{\rm G}$
   \citep{narayan95}, which gives a transition radius $R_{\rm tr}
   \simeq 1.5 \times 10^3 r_{\rm G}$. In the evaporation
    model \citep{meyer00,liubf02}, 
\begin{equation}
  R_{\rm ev} \simeq 18.3 \dot{m}^{-0.85} r_{\rm G} ,
  \label{eqa2strans}
\end{equation}
   which gives a consistent transitional radius $R_{\rm tr} = R_{\rm
   ev} \simeq 1.2 \times 10^3 \, r_{\rm G}$. 

    \citet{rezzolla03b} and \citet{zanotti03} showed that if it 
    is globally perturbed,  a torus or a thick disk of finite 
    radial extent oscillates  globally. The global oscillations lead
    to quasi-periodic accretion with frequencies of harmonic
    relationship 1:2:3:4$\dots$ \citep{zanotti03}
\begin{equation}
  f_{\rm m}= m f_0 ,
\end{equation}
    where $m=1$, 2, 3, $\dots$. The fundamental oscillation
    frequency $f_0$ is given with $ f_0 = c_{\rm s} / L$ where $c_{\rm 
    s}$ is sound speed and $L = p/(dp/dr)$ is the radial length scale
    of pressure {\it p}. Note that $f_{\rm
    m}$ in \citet{zanotti03} relates to the thick disk
    eigen-frequencies $f_{\rm n} = [(2+n)/2] f_0$ for $n = 0$, 2, 
    4, $\dots$ and $m=(2+n)/2$ \citep{rezzolla03b}. Therefore, the
    trapped acoustic wave should have zero initial phase difference. The
    quasi-periodic p-mode oscillations of a small-size torus orbiting
    around a black hole have been suggested to explain the
    high-frequency QPOs (HFQPOs) in black hole X-ray binary systems
    \citep{rezzolla03}, which is observed to be in ratios of small
    integers \citep[e.g.][]{abram01}. If the inner thick accretion
    flow of finite radial size in AO 0235+164 is perturbed by global
    perturbation or by strong local periodic agent, the accretion rate
    would vary 
    quasi-periodically, triggering quasi-periodical plasma injection
    into relativistic jet due to jet-disk coupling and leading to the
    quasi-periodical variation of jet emission. As it describes an
    ADAF quite well except at two boundaries, the self-similar
    solution of ADAF is used in estimation of the fundamental
    oscillation frequency $f_0$. From the self-similar solution for
    the ratio of specific 
    heats $\gamma = 5/3$ \citep{narayan94,narayan00}, the pressure $p
    \propto \rho c_{\rm s}^2 \propto R^{-5/2}$ and sound speed $c_{\rm
    s} = {\sqrt{10} \over 3 \alpha} \left[\left(1 + {18 \alpha^2 \over
    25}\right)^{1/2} -1\right]^{1/2} v_{\rm K} \simeq (2/5)^{1/2}
    v_{\rm K}$ for $\alpha^2 \ll 1$, where $v_{\rm K}$ is the Keplerian
    velocity at radius $R$. Then the fundamental frequency is
\begin{eqnarray}
  f_0 & = & c_{\rm s} / L \simeq (5/2)^{1/2} \Omega_{\rm K} \cr
  &\simeq& 1.13 \, {\rm yr}^{-1} \left({R \over 10^3 r_{\rm
      G}}\right)^{-3/2} \left({M_{\rm BH} \over 10^8
    M_\odot}\right)^{-1} , 
\end{eqnarray}
    where $\Omega_{\rm K}$ is the Keplerian angular frequency and
    $r_{\rm G} = 2 GM /c^2 = 2.95 \times 10^{13} \, {\rm cm} \left(M /
    10^8 M_\odot\right)$ is the Schwarzschild radius. The 
    fundamental oscillation frequency of the thick ADAF of
    AO 0235+164 is $f_0 \simeq 0.182 \, {\rm yr}^{-1} \left(R_{\rm tr}
    /1.2\times 10^3 r_{\rm G}\right)^{-3/2} \left({M_{\rm BH}
    /4.72\times 10^8 M_\odot}\right)^{-1}$ and the
    fundamental period is $P_0 = 5.48 \, {\rm yr} \left(R_{\rm tr}
    /1.2\times 10^3 r_{\rm G}\right)^{3/2} \left({M_{\rm BH}
    /4.72\times 10^8 M_\odot}\right)$. The oscillation model for torus
    or thick disk predicts a series of overtones in a sequence
    1:2:3:$\dots$, independent of the disk radial extent and of the
    variation of sound speed with radius. The fundamental frequency
    and all the overtones are predicted to have nearly zero initial
    phase difference.  

    With the redshift $z = 0.94$ of AO 0235+164, we have the expected  
    lowest frequency $f_{\rm obs} = f_0 /(1+z) \simeq 
     2.57 \times 10^{-4} \, {\rm
    days}^{-1} \left(R_{\rm tr} /1.2\times 10^3 r_{\rm G}\right)^{-3/2}
    \left({M_{\rm BH} /4.72\times 10^8 M_\odot}\right)^{-1}$ 
    and period $P_{\rm obs} \simeq 10.7 \, {\rm yr} \left(R_{\rm tr}
    /1.2\times 10^3 r_{\rm G}\right)^{3/2} \left({M_{\rm BH}
    /4.72\times 10^8 M_\odot}\right)$,
    respectively, which are consistent with the detected frequency
    $\nu_1 \simeq (2.28 \pm 1.10) \times 10^{-4} \, {\rm days}^{-1}$
    and period $P_1 \simeq (12.0 \pm 5.8) \, {\rm yrs}$ (or $2P_2 =
    10.9 \, {\rm yr}$ at higher accuracy) at 8 GHz and
    other radio frequencies. As the observed lowest frequency, the
    harmonics and the difference of initial phases are consistent with
    the predictions of a thick ADAF oscillation model,  we suggest
    that the multiple harmonic QPOs in the radio light curves of AO
    0235+164 are due to the perturbed p-mode oscillations of an ADAF
    with a finite radial extent $R_{\rm tr} \simeq 1.2 \times 10^3
    r_{\rm G}$ between a central supermassive black hole and an outer
    cool thin disk.  
    
    There are two difficulties needed to be addressed with the
    model. The first one is the relative strength of QPOs in the
    power spectrum. In the theoretical power spectrum of oscillation
    accretion, the fundamental QPO is the strongest and the strength
    of harmonics decreases with the oscillation 
    frequency \citep{rezzolla03b,zanotti03}.  The Lomb power
    spectrum periodic analyses show that the $rms$ of QPOs indeed
    decreases with frequency except the 6th harmonics with
    frequency $\nu_6 = (1.511 \pm 0.099) \times 10^{-3} \, {\rm
    day}^{-1}$ or period $P_6 = 1.81 \pm 0.12 \, {\rm yr}$, which is
    exceptionally prominent and inconsistent with the prediction of
    the 
    disk oscillation model. The other is on the nature of perturbation
    source. There is no obvious global or local perturbation source in
    a SMBH-thick disk system in AO 0235+164. We will discuss the two
    difficulties in Sec.~\ref{sec:pertbinary}.

%------------------------------
\section{Perturbation sources and supermassive black hole binary}
\label{sec:pertbinary}

    Although in Sec.~\ref{sec:diskosc} we suggest that the observed
   multiple harmonic QPOs in AO 0235+164 would be due to the p-mode
   oscillations of a thick accretion disk (ADAF or RIAF), the nature
   of strong excitation mechanisms, either local
   \citep{rubio05a,rubio05b} or global
   \citep{zanotti03,rezzolla03}, is still an open question, in
   particularly for a system of a thick disk with large radial
   extent around a SMBH. Here we suggest that a SMBHB embedding in
   the thick accretion disk would serve as such a local, strong and
   periodic perturbation source as required by the p-mode
   oscillations of a thick accretion disk investigated by 
   \citet{rubio05b}. A SMBHB may form in galaxy minor
   merger and is introduced in a helical jet model for AO 0235+164
   \citep{osto04} and later in explaining the rapid changes in the
   VLBI (Very Long Baseline Interferometry) jet position
   angles of the radio source \citep{frey06}. A SMBHB embedding in an 
   ADAF or RIAF cannot open a gap in an accretion disk, because to
   open a gap the disk opening angle $H/R$ should be 
   $H/R \la 0.3$ and the mass ratio {\it q} of the secondary and the 
   primary black holes should be $q \leq 1$ and $q >
   q_{\rm  min} = (81 \pi / 8) \alpha (H / R)^2 \ga 3$ for an 
   ADAF with $H/R \sim 1$ and $0.1 < \alpha < 0.3$
   \citep{lin86}.  The 
   interaction with an RIAF is hydro-dynamically unimportant to the
   evolution of a SMBHB \citep{liu04}, but the gravitational 
   attraction of the secondary to the plasma in the thick disk changes
   periodically and triggers a series of density waves which propagate
   inwards and outwards in the thick disk with a limited radial extent. When
   they arrive at the inner or the outer boundaries, the acoustic waves
   are reflected backwards with a very large fraction of the incoming 
    energy, if the acoustic frequency is about the fundamental
   frequency or harmonics of the cavity. Other acoustic waves 
   are almost completely be absorbed by the boundaries. This kind of p-mode
   oscillations in a tours or thick accretion disk were first
   discussed for global perturbation by \citet{rezzolla03b} and later
   for local strong periodic agent by \citet{rubio05a}. A SMBHB thus
   perfectly serves as a local strong perturbation agent to a
   SMBH-thick disk system. 

   For a SMBHB with separation $a \la R_{\rm tr}$, the orbital period
   $P_{\rm b}$ should be 
\begin{eqnarray}
  P_{\rm b} & = & 2 \pi \left[{a^3 \over G M_{\rm BH} \left(1+
      q\right)}\right]^{1/2} \cr
      & \la & 8.8 \, {\rm yr} \, \left(R_{\rm tr} \over 10^3 r_{\rm 
      G}\right)^{3/2} \left({M_{\rm BH} \over 10^8 M_\odot}\right)
      \left(1+ q\right)^{-1/2} .
\label{eq:obplimit}
\end{eqnarray}
   In the observer's frame, the period is observed to be $P_{\rm obs} =
   (1+z) P_{\rm b} \la 80 \, {\rm yr}$, which may  or may not be
   within the detection window of the three databases which we
   used. \citet{rubio05b} showed that in addition to 
   the strong QPO with very high quality factor $Q$ at the
   perturbation period $P_{\rm b}$, the oscillations of a perturbed
   thick disk should be detected significantly at $P_{\rm b} / 2$ and
   possibly also at $2 P_{\rm b}$. In the helical-jet model
   \citep{osto04}, the 
   SMBHB orbital period is identified with $P_{\rm obs}=P_2 \sim 5 - 6
   \, {\rm yrs}$. If the helical jet model is correct, the SMBHB has a
   separation $a \simeq 166 r_{\rm G}$, which is much smaller than the
   transitional radius $R_{\rm tr}$ and implies an embedding SMBHB
   in ADAF (or RIAF). In this scenario, SMBHB should excite thick disk
   oscillations not only at period $P_2$ but also at periods $P_4$ and
   probably $P_1$. One difficulty with the explanation is that one has to
   explain the exceptional prominence of the 6th QPO with very high
   quality factor $Q$, which should be the weakest among six QPOs
   and may not be detectable in the disk oscillation model. Therefore,
   a more reasonable suggestion would be that the 6th QPO of period
   $P_6 = 1.81 \, {\rm yr}$ corresponds to the SMBHB orbital
   motion. In this scenario, a QPO at period $P_6 / 2 = 0.91 \, {\rm
   yrs}$ should also be significant, which might have already been
   detected with $P= 1.03 \pm 0.03 \, {\rm yrs}$ at the Lomb power
   spectrum at 8 GHz in Fig.~\ref{0235l8c}. However, a QPO with a
   period about one year may not be identified correctly due to the
   one-year astronomical cycle. 

   For a SMBHB with an orbital period $P_{\rm obs} =1.81 \, {\rm yr}$
   or $5.46 \, {\rm yr}$, the binary separation is 
\begin{eqnarray}
  a & = & r_{\rm G} \left[{P_{\rm obs} \over \left(1+z\right) 2
   \sqrt{2} \pi}  
   {c \over r_{\rm G}}\right]^{2/3} \left(1 + q\right)^{1/3} \cr
   & \simeq & 166 r_{\rm G} \left(1 + q\right)^{1/3} \quad {\rm if\ 
   P_{obs} =5.46 \, yr, \ or } \cr
   & \simeq & 79 r_{\rm G} \left(1 + q\right)^{1/3} \quad {\rm if\ 
   P_{obs} =1.81 \, yr} .
\label{eqbinsep}
\end{eqnarray}
   Because the interaction with an ADAF or RIAF is negligible and the
   evolution of SMBHB is dominated by the gravitational wave radiation
   \citep{liu04}, the lifetime of SMBHB with a circular orbit in AO
   0235+164 due to the gravitational wave radiation is \citep{peters63} 
\begin{eqnarray}
  t_{\rm gw} & =& {a \over |\dot{a}|} =  {5 \over 8} \left({a \over
    r_{\rm G}}\right)^4 q^{-1} \left(1 + q\right)^{-1} 
  {r_{\rm G} \over c}  \cr
  &\simeq & 7.0 \times 10^6 \, {\rm yr} \, \left({q \over
    0.01}\right)^{-1} \left(1 + q\right)^{1/3} \quad {\rm if\ 
    P_{obs} =5.46 \, yr, \ or }  \cr
  & \simeq & 3.6 \times 10^5 \, {\rm yr} \, \left({q \over
    0.01}\right)^{-1} \left(1 + q\right)^{1/3} \quad {\rm if\ 
   P_{obs} =1.81 \, yr} .
\label{eq:gwlife}
\end{eqnarray}
    Both identifications of the SMBHB orbital motion suggest a
    short-lived binary. However, it is also possible that the
    separation of SMBHB is 
    $a\sim 10^3 r_{\rm G}$ and the orbital period $P_{\rm b} \sim
    70 \, {\rm yr}$ is out of the observational window. A SMBHB
    with such a large separation is long-lived with $t_{\rm gw} \simeq
    9.2 \times 10^9 \, {\rm yr}\, \left({q \over 0.01}\right)^{-1}
    \left(1 + 
    q\right)^{1/3}$. The observations of the changes of VLBI jet
    positional angle may help to resolve the puzzles.

%------------------
\section{Discussions and conclusions}
\label{concl}

    We have investigated the radio variabilities of AO 0235+164 and 
    analyzed the periods of radio light curves at 4.8, 8, 14.5, 22,
    and 37 GHz, basing 
    on the databases of UMRAO, NRAO and Mets\"{a}hovi
    Observatory. We first study the consistence of the UMRAO's and
    NRAO's data at 8 GHz for AO
    0235+164 and then construct a combined radio light curve with the
    observational data from the two databases. The periodic
    analyses with three kinds of classical periodic analysis
    methods show with unprecedentedly high signal-to-noise ratio that
    the variations of the combined radio light curves are composed of
    six periodic variations plus random fluctuations. In the 
    power spectrum, the peaks are Gaussian and identified with QPOs of
    periods $P_1 = 12.02 \, {\rm yr}$, $P_2 = 5.45 \, {\rm yr}$, $P_3
    = 3.62 \, {\rm yr}$, $P_4 = 2.83 \, {\rm yr}$, $P_5 = 2.15 \, {\rm
    yr}$, and $P_6 = 1.81 \, {\rm yr}$. The coherent quality factor
    $Q$ of the six QPOs are limited by the monitoring program
    time. We discover a harmonic relationship for the six QPOs,
    based on the ratio of QPO frequencies in a sequence of small
    integers 1:2:3:4:5:6 and on the zero or $\pi$ difference of
    initial phases of QPOs. The harmonic relationship  
    suggests that the six QPOs should have the same physical origin. 
    Among the six harmonic QPOs, the second one with period $P_{\rm 2}
    \simeq 5.46 \pm 0.47 \, {\rm yrs}$ is the strongest and is
    identified with the fundamental period. The five strong harmonics 
    reported in some literature are confirmed here with an
    unprecedentedly high signal-to-noise and accuracy, while the
    weakest QPO of period $P_5 = 2.15 \, {\rm yrs}$ is reported first
    time in this paper. 

    We investigate the dependence of harmonic relationship, coherent
    quality factor, the relative {\it rms}, the differences of the
    initial phase of QPOs on radio frequency. The results show 
    that all the QPO properties are independent of radio
    frequencies. The six harmonic QPOs give a good fit to the outburst
    structures of the radio light curves at 4.8, 8, 14.5, 22 and 37
    GHz, and predict major outbursts and outburst structures
    significantly different from that given with a single period of $P
    \simeq 5.46 \, {\rm yr}$ \citep{rai01}. Both the single-period and
    the multiple harmonic QPO (for the low frequencies) scenarios
    predict a major outburst in the early half of 2004, while the 
    combinations of six harmonic QPOs suggest only minor
    fluctuations during that period but a major burst in the late half
    of 2004 or the early half 2005 at 37 GHz, as it is shown in
    Fig.~\ref{fit8}. In our periodic analyses of radio light curves at
    22 GHz and 37 GHz, we have used the data after 2001 and the
    harmonic QPO scenario predict a major outburst in the early
    2005. At lower frequencies we cannot get the observational
    data in recent years and make the analyses basing on the data
    before 2001. Therefore, the predictions are slightly different
    for different wave-bands and the predicted major bursts may be in
    the early 2005. However, the observations of a WEBT (the Whole
    Earth Blazar Telescope) campaign
    started from 2003 show that AO 0235+164 is quiet in recent years
    and the predicted major outbursts in the spring of 2004 did not
    come \citep{rai05}. Is the predicted major outburst just delayed 
    significantly or is the activity of the object interrupted due to
    some reasons? When will the object re-activate with a different
    active cycle? To answer the questions, more observations for
    another several years are needed. 

    Either a significant delay of outbursts or the interruption of
    activity for some time would exclude the single period scenario
    and the simple version of a helical jet model but may be
    consistent with the SMBHB-thick disk oscillation model. 
    In a SMBHB-thick disk oscillation scenario, the multiple QPOs
    in the radio light curves of AO 0235+164 are due to quasi-periodic 
    injection of plasma from an oscillation thick disk of a finite radial
    extent into
    relativistic jet. A torus or a thick accretion disk of finite
    radial extent oscillates and accretes matter quasi-periodically
    with small integer ratio of frequencies with 1:2:3:$\dots$, if it
    is perturbed by a global perturbation or a local strong periodic
    agent \citep{zanotti03,rezzolla03,rubio05b}. The fundamental
    oscillation frequency of a thick disk is determined by the radial
    extent and sound speed. We estimated the central black hole mass
    $M_{\rm BH} = (4.72\pm 2.04) \times 10^8 M_\odot$ and the
    dimensionless 
    accretion rate $\dot{m} \simeq 0.007$ of AO 0235+164, based on
    the observations of emission line properties. The estimated
    dimensionless accretion rate with the current knowledge of
    accretion around black hole suggests that the inner region of
    accretion disk in AO 0235+164 is a geometrically thick disk
    (i.e. ADAF or RIAF) with radial extent about $1.2 \times 10^3
    r_{\rm G}$ and the outer region is a truncated cool thin disk.
    With the self-similar solution of ADAF, we analytically estimate
    the fundamental oscillation frequency $f_{\rm obs} \simeq 0.0938
    \, {\rm yr}^{-1}$ or period $P_{\rm obs} = 10.7 \,{\rm yr}$ in
    the observer's frame, which together with the predicted overtones
    are consistent with the observed multiple QPOs with lowest
    frequency $P_1 \simeq 12.02 \pm 5.79 \, {\rm yr}$ (or $2P_2 = 10.9
    \, {\rm yrs}$ at higher accuracy) and harmonics. As there is no
    natural global or strong local perturbation source in standard
    SMBHB-accretion disk system to excite the global
    oscillations of a thick disk of a large radial extent $R_{\rm d}
    \sim 10^3 r_{\rm G}$, we suggest a SMBHB rotating within the thick
    disk to excite the p-mode oscillations of RIAF in AO 0235+164. The
    rotation of the secondary triggers acoustic waves in an ADAF (or
    RIAF), which propagate inwards and outwards in the thick disk and
    are reflected backwards by the inner and outer 
    boundaries. The p-mode oscillations with frequencies of about the
    fundamental oscillation frequency and the overtones are trapped by
    the cavity. The orbital motion of a SMBHB could lead to the
    formation of a helical morphology of relativistic jet and to the 
    rapid changes of VLBI jet position angles. Therefore, in the
    SMBHB-disk oscillation model, the relativistic jet can be also
    helical. The orbital motion may correspond to the 6th QPO with
    $P_6 = 1.81 \, {\rm yrs}$, the 2nd QPO with $P_2 = 5.46 \, {\rm
    yrs}$, or have not been observed. Long term VLBI monitoring of
    rapid changes of jet positional angles would help to answer
    the question. One of the differences with the simple helical jet
    model is that in the SMBH-thick disk oscillation scenario the
    minor burst events in the radio light curves is no longer random
    but together with the major outbursts are the combinations
    of harmonic six QPOs which are the p-mode oscillations of the RIAF
    and the binary orbital motion. This is a first time report of
    multiple harmonic QPOs and the possible detection of perturbed
    disk p-mode oscillations in AGNs. 

    The second difference of predictions of the helical jet model and
    the SMBHB-thick disk oscillation scenario is that periodic
    outbursts in the former model are persistent and regular while in
    the later scenario are temporary with moderate coherence quality
    factor $Q$ and depend on the dimensionless accretion rate. In the
    SMBHB-thick oscillation scenario, resonant harmonic QPOs would
    disappear from the radio light curves or the periodic activities
    of the radio source are interrupted, when the accretion rate
    increases only about two times aand becomes larger than the
    critical accretion rate $\dot{m}_{\rm 
    cr}$ and the inner region of accretion disk becomes a
    geometrically thin cool disk. Therefore, the quiescence and the
    absence of outbursts from the radio light curves of AO 0235+164
    may be because of the transition of accretion mode from a RIAF to
    a standard thin cool accretion disk due to the moderate increase of
    accretion rate. The X-ray spectral observations with the Chandra
    satellite in August 2000 and with the XMM-Newton satellite in
    August 2004 show tentative detections of the red-shifted Fe
    K$\alpha$ fluorescent emission line in AO 0235+164 \citep{rai06},
    which suggests that the accretion disk at the immediate vicinity
    of the SMBH is very likely a standard thin accretion disk
    \citep{fabian00}. Therefore, the absence of outbursts from the 
    radio light curves and the interruption of the harmonic QPOs can
    be understood in the SMBHB-thick disk oscillation scenario as the
    transition of accretion mode due the increase of
    accretion rate. In a jet-disk coupling scenario, an object 
    becomes radio quiet when the accretion disk becomes a standard
    thin disk \citep{fend04}.

    We notice that in the helical jet model, the separation of SMBHB 
    is much smaller than the transitional radius $R_{\rm tr}$ of inner
    thick RIAF and outer thin disk. Therefore, the SMBHB in the helical
    jet model for AO 0235+164 should also excite the p-mode
    oscillations of an RIAF, leading to quasi-periodic accretion and
    quasi-periodic injection of plasma into the relativistic 
    jet. Therefore, multiple QPOs with harmonic relationship are also
    expected by the helical jet model for AO 0235+164, if we consider
    the RIAF p-mode oscillations. 

    The embedding secondary black hole 
    spherically accretes matter from the RIAF and radiates energy in
    X-ray. The accretion radius of the secondary black hole is $r_{\rm
    acc} \approx {Gm \over c_{\rm s}^2 + (\Delta{v})^2}$, where
    $\Delta v$ is the differential velocity of the secondary black
    hole and the plasma in the thick disk. For a minor merger with 
    $q \ll 1$, we obtain $r_{\rm acc} \ll R$ and $(\Delta{v})^2 \ll
    c_{\rm s}^2$. From the self-similar solution $c_{\rm s}^2 = (2/5)
    V_{\rm K}^2$ for $\gamma = 5/3$ \citep{narayan94,narayan00}, we
    have $r_{\rm acc} \approx (5/2) q (1+q) R$. The accretion rate of
    the secondary black hole is  
\begin{equation}
  \dot{M}_{\rm s} \sim 4 \pi r_{\rm acc}^2 c_{\rm s} \rho ,
\label{eq:secacc}
\end{equation} 
    where the mass density $\rho$ is given with the self-similar
    solution and $\rho = {5 \sqrt{5} \over 12 \sqrt{2} \pi} \dot{M}_{\rm
    Edd} \dot{m} \alpha^{-1} R^{-2} V_{\rm K}^{-1}$. From
    Eq.~(\ref{eq:secacc}), we have $\dot{M}_{\rm s} \sim {25 \sqrt{5}
    \over 6 \sqrt{2}} q^2 (1+q)^2 \dot{M}_{\rm Edd} \dot{m}
    \alpha^{-1}$. Defining the Eddington accretion rate for the
    secondary black hole 
    $\dot{M}_{\rm Edd}^s = L_{\rm Edd}^s / 0.1 c^2$ with $ L_{\rm
    Edd}^s = 1.26 \times 10^{38} (m / M_\odot) \, {\rm ergs\;
    s^{-1}}$, we have the dimensionless accretion rate of the
    secondary black hole, $\dot{m}_{\rm s} \equiv \dot{M}_{\rm s} /
    \dot{M}_{\rm Edd}^s \sim 6.6 (1+q)^2 (q/\alpha) \dot{m}$. For the
    typical parameters 
    $q = 0.01$ and $\alpha = 0.13$, the relative accretion rate of the
    secondary is $\dot{m}_{\rm s} \sim 0.51 \dot{m} \sim 5.1 \times
    10^{-3}$, while $\dot{m}_{\rm s} \sim 5.1 \times 10^{-2}$ for
    $q=0.1$. The X-ray emission from the accretion of the secondary
    black hole is small and may not be observable in AO 0235+164, 
    because of the strong X-ray radiations both from the relativistic 
    jet and from accretion disk around the primary. A different
    $\gamma$ value does not change the conclusion. 

    Our work implies that if a thick accretion disk with finite radial
    extent is perturbed globally or locally by, e.g. SMBHB, the
    p-mode oscillations can be observed as multiple harmonic QPOs,
    when the relativistic jet is along the line of
    sight. Periods have been reported and SMBHBs have been
    suggested in 
    literature for many AGNs. So, multiple QPOs with a
    harmonic relationship of frequencies are expected by the
    SMBHB-disk oscillation model in moderately
    low luminosity AGNs with SMBHB at center, which may have
    accretion configuration like the one in the BL Lac object AO
    0235+164. We will report our periodic analysis results on a large
    sample of AGNs in a coming paper \citep{zhao05}.

\acknowledgments

    We are grateful to L. Rezzolla, J. Miller,  H. Su, R. Wang,
    X. Chen, Y.B. Liu, and B.F. Liu for many insightful discussions
    and help. Many thanks are due to the referee for his/her
    constructive comments, which help us to improve the presentations
    significantly. We 
    thank Kari Nillsson for helpful explanations on the spectroscopic 
    data of AO 0235+164. This work is supported by the National
    Science Foundation of China (No. 10203001, No. 10573001,
    No. 10473001, No. 10525313) and partly by the National Key Project on 
    Fundamental Researches (TG 1999075403). This research has made use
    of:  

--the data from the University of Michigan Radio Astronomy
  Observatory, which is supported by the National Science Foundation
  and by funds from the University of Michigan; and 

--the data from the Green Bank Interferometer (GBI), National Radio
  Astronomy Observatory (NRAO). The Green Bank Interferometer is a
  facility of the National Science Foundation operated by the NRAO in
  support of NASA High Energy Astrophysics programs.

\clearpage

%-------
\begin{table*}
\tabletypesize{\small}
\caption{Hormanic QPOs in the radio light curves at 4.8 GHz, 8 GHz,
  14.5 GHz, 22 GHz and 37 GHz analyzed with the Lomb power spectrum
  method. 
\label{idf}}

\begin{tabular}{ c| r@{$\pm$}l r@{$\pm$}l r@{$\pm$}l r@{.}l c r@{.}l c c r@{.}l }
\multicolumn{16}{c}{ }\\\hline\hline

$\nu_{obs}$ & \multicolumn{2}{c}{$\nu$}                   &
\multicolumn{2}{c}{\it rms}  & \multicolumn{2}{c}{P}    & \multicolumn{2}{c}{Q} & $\rm\chi^2 $ &\multicolumn{2}{c}{N} & $\nu/\nu_s$ & Id.       & \multicolumn{2}{c}{$\rm\Delta\phi_0$}\\

$\bar{P}_s$ ($\rm yr^{-1}$) & \multicolumn{2}{c}{$(10^{-3}\rm
  day^{-1})$} & \multicolumn{2}{c}{(\%)} & \multicolumn{2}{c}{(yr)} &
\multicolumn{2}{c}{ } &              &\multicolumn{2}{c}{ } &
&   & \multicolumn{2}{c}{($\pi$)} \\\hline

8 GHz       &  0.2279&0.1098  &  7.29&0.35  & 12.02&5.79  &  1&76  &  0.431  &  2&2  &  0.45  &  1  & -0&39  \\
            &  0.5029&0.0787  &  9.55&0.27  &  5.45&0.85  &  5&43  &  0.485  &  4&8  &  1.00  &  2  &  0&00  \\
  5.46      &  0.7568&0.1105  &  8.04&\ 0.44  &  3.62&0.53  &  5&82  &  0.485  &  7&3  &  1.50  &  3  & -0&17  \\
 $\pm$0.47  &  0.9691&0.0806  &  6.25&0.41  &  2.83&0.24  &  10&21  &  0.485  &  9&3  &  1.93  &  4  &  0&98  \\
            &  1.275 &0.074  &  4.70&0.54  &  2.15&0.12  & 14&69  &  0.162  & 12&2  &  2.54  &  5  & -0&58  \\
            &  1.511 &0.099  &  7.13&0.36  &  1.81&0.12  & 12&98  &  0.332  & 14&5  &  3.00  &  6  &  0&07  \\\hline

4.8 GHz            &  0.2568&0.0797  & 10.86&0.66  & 10.67&3.31  &  2&74  &  0.132  &  2&3  &  0.54  &  1  & -0&35  \\
                   &  0.4716&0.0700  &  9.02&0.79  &  5.81&0.86  &  5&72  &  0.031  &  4&2  &  1.00  &  2  &  0&00  \\
    5.62    &  0.7254&0.0771  &  8.20&0.87  &  3.78&0.40  &  7&99  &  0.085  &  6&5  &  1.54  &  3  & -0&19  \\
  $\pm$0.44  &  0.9295&0.0780  &  7.08&1.01  &  2.95&0.25  &  10&12  &  0.019  &  8&5  &  1.97  &  4  &  1&19  \\
            &    \multicolumn{12}{c}{ }                                                              & (5) &(-0&41) \\
            &  1.522 &0.066  &  8.27&0.86  &  1.80&0.08  & 19&53  &  0.006  & 13&6  &  3.23  &  6  &  0&36  \\\hline

14.5 GHz    &  0.2019&0.0558  &  6.19&0.57  & 13.57&3.75  &  3&07  &  0.014  &  1&9  &  0.42  &  1  & -0&34  \\
            &  0.4826&0.0532  &  8.10&0.44  &  5.68&0.63  &  7&70  &  0.251  &  4&4  &  1.00  &  2  &  0&00  \\
    5.53    &  0.7299&0.0516  &  6.21&0.57  &  3.75&0.27  & 12&01  &  0.033  &  6&7  &  1.51  &  3  & -0&11  \\
 $\pm$0.33  &  0.9776&0.0568  &  7.58&0.47  &  2.80&0.16  & 14&62  &  0.041  &  9&0  &  2.03  &  4  &  1&22  \\
            &  1.297 &0.060   &  4.95&0.72  &  2.11&0.10  & 18&36  &  0.002  & 11&9  &  2.69  &  5  & -0&40  \\
            &  1.472 &0.067   &  6.72&0.53  &  1.86&0.08  & 18&77  &  0.004  & 13&5  &  3.05  &  6  &  0&44  \\\hline

22 GHz      &  0.3138&0.0625  &  7.73&0.23  &  8.73&1.74  &  4&26  &  0.087  &  2&6  &  0.66  &  1  & -0&12  \\
            &  0.4755&0.0924  &  8.96&0.20  &  5.76&1.12  &  4&37  &  0.087  &  3&9  &  1.00  &  2  &  0&00  \\
   5.55     &  0.6890&0.0701  &  7.47&0.23  &  3.98&0.40  &  8&35  &  0.025  &  5&6  &  1.45  &  3  & -0&34  \\
 $\pm$0.47  &  1.027 &0.079   &  6.06&0.29  &  2.67&0.21  &  11&04  &  0.015  &  8&4  &  2.16  &  4  &  0&94  \\
            &   \multicolumn{12}{c}{ }      & (5) &(-0&29) \\
            &  1.516 &0.078   &  6.56&0.27  &  1.81&0.09  & 16&59  &  0.015  & 12&4  &  3.19  &  6  & -0&02  \\\hline

37 GHz      &  0.1214&0.0753  &  5.99&0.27  & 22.57&14.00  &  1&37  &  0.039  &  1&03 &  0.30  &  1  & -0&55  \\
            &  0.4065&0.0948  &  7.07&0.23  &  6.74&1.57  &  3&64  &  0.040  &  3&5  &  1.00  &  2  &  0&00  \\
  5.77      &  0.7376&0.0644  &  3.95&0.41  &  3.71&0.32  &  9&73  &  0.040  &  6&3  &  1.81  &  3  &  0&19  \\
  $\pm$0.64 &  0.9879&0.1305  &  4.75&0.41  &  2.78&0.37  &  10&74  &  0.073  &  8&4  &  2.43  &  4  & -0&42  \\
            &    \multicolumn{12}{c}{ }                                                              & (5) &( 0&36) \\
        &  1.432 &0.113   &  5.93&0.27  &  1.91&0.15  &  9&13  &  0.073  & 12&2  &  3.52  &  6  & -0&77  \\\hline

\end{tabular}

\end{table*}

\end{document}